\documentclass{aa}
\usepackage{lscape}
\usepackage{graphicx}
\usepackage{txfonts}
\usepackage{natbib}
\usepackage{multirow}
\usepackage[usenames, dvipsnames]{color}

\bibpunct{(}{)}{;}{a}{}{,} 
\begin{document}

   \title{Electrical properties and porosity of the first meter of the nucleus of 67P/Churyumov-Gerasimenko }
\subtitle{As constrained by the Permittivity Probe SESAME-PP/Philae/Rosetta}
   \author{Anthony Lethuillier
        \inst{1}
         \and
         Alice Le Gall
        \inst{1}
         \and
        Michel Hamelin 
        \inst{2}
         \and
        Walter Schmidt
        \inst{3}
         \and
        Klaus J. Seidensticker
        \inst{4}
         \and
        Réjean Grard
        \inst{5}
         \and
        Valérie Ciarletti
        \inst{1}
         \and
        Sylvain Caujolle-Bert
        \inst{1}
         \and
        Hans-Herbert Fischer
        \inst{6}
         \and
        Roland Trautner
        \inst{7}
          }

   \institute{LATMOS/IPSL, UVSQ Universités Paris-Saclay, UPMC Univ. Paris 06, CNRS, Guyancourt, France \email{anthony.lethuillier@latmos.ipsl.fr}
                                                        \and
                                                        LATMOS/IPSL, UPMC Univ. Paris 06 Sorbonne Universités, UVSQ, CNRS, Paris, France 
                                                        \and
                                                        Finnish Meteorological Institute, Helsinki, Finland
                                                        \and
                                                        Deutsches Zentrum für Luft- und Raumfahrt, Institut für Planetenforschung, Rutherfordstraße 2, 12489 Berlin, Germany
                                                        \and
                                                        Retired from ESA/ESTEC, 11 rue des Forges, Paunay, 79400, Saivres, France
                                                        \and
                                                        Deutsches Zentrum für Luft- und Raumfahrt, Raumflugbetrieb und Astronautentraining, MUSC, Linder Höhe, 51147 Köln, Germany
                                                        \and
                                                        ESA/ESTEC, Noordwijk, The Netherlands
                                                        }
    \date{Received 15 February 2016 / Accepted 01 April 2016}

  \abstract
   {Comets are primitive objects, remnants of the volatile-rich planetesimals from which the solar system condensed. Knowing their structure and composition is thus crucial for the understanding of our origins. After the successful landing of Philae on the nucleus of 67P/Churyumov-Gerasimenko in November 2014, for the first time, the Rosetta mission provided the opportunity to measure the low frequency electrical properties of a cometary mantle with the permittivity probe SESAME-PP (Surface Electric Sounding and Acoustic Monitoring Experiment - Permittivity Probe).}
   {In this paper, we conduct an in-depth analysis of the data from active measurements collected by SESAME-PP at Abydos, which is the final landing site of Philae, to constrain the porosity and, to a lesser extent, the composition of the surface material down to a depth of about 1 meter.}
   {SESAME-PP observations on the surface are then analyzed by comparison with data acquired during the descent toward the nucleus and with numerical simulations that explore different possible attitudes and environments of Philae at Abydos using a method called the Capacity-Influence Matrix Method.}
   {Reasonably assuming that the two receiving electrode channels have not drifted with respect to each other during the ten-year journey of the Rosetta probe to the comet, we constrain the dielectric constant of the first meter below the surface at Abydos to be $>2.45\pm0.20$, which is consistent with a porosity $<50\%$ if the dust phase is analogous to carbonaceous chondrites and $<75\%$ in the case of less primitive ordinary chondrites. This indicates that the near surface of the nucleus of 67P/Churyumov-Gerasimenko nucleus is more compacted than its interior and suggests that it could consist of a sintered dust-ice layer.}
   {}


\maketitle
%

\section{Introduction}

The European Space Agency (ESA) Rosetta spacecraft \citep{Glassmeier2007} reached its final target in summer 2014: the comet 67P/Churyumov-Gerasimenko (hereafter 67P/C-G). On November 12, 2014, the Philae probe \citep{Bibring2007} landed on the surface of the nucleus at a distance of $2.99$\, au from the Sun. Among the instruments onboard Philae, the Permittivity Probe instrument (SESAME-PP), which is part of the Surface Electric Sounding and Acoustic Monitoring Experiment (SESAME) instrument package \citep{Seidensticker2007} operated both during descent and on the surface. The primary scientific objective of this experiment is to measure the low-frequency complex permittivity, i.e., the dielectric constant and electrical conductivity, of the first meter below the surface of the cometary nucleus. These measurements provide a unique insight into the composition and, in particular, the water content and porosity of the comet mantle.

Comets are primitive objects with a composition that has barely changed in $4.6$\,Gyr. As such, they likely hold important clues on the youth and evolution of the solar system. Comets may have also brought to our planet Earth organic molecules and most of the water that sustains life, although the Deuterium to Hydrogen ratio discrepancy between 67P/C-G and Earth oceans indicates that at least the Jupiter family comets may not be the major contributing source of water \citep{Altwegg14}. Knowing their internal structure and composition is thus crucial for the understanding of our origins. In particular, little is known about the composition of cometary mantles. Prior to the arrival of Rosetta at 67P/C-G, a variety of models were proposed including the icy-glue model \citep{Gombosi1986}, the icy conglomerate model \citep{Klinger1985}, the fluffy aggregate model \citep{Donn1989}, and the primordial rubble model \citep{Weissman1986}; these models are mainly based on observations collected during the flybys of comet 1P/Halley in the mid-eighties. In these models, as in the most recent layered pile model \citep{Belton2007}, the mantle generally consists of residues that remain on the surface after the sublimation of volatiles. This deposition layer may vary in size and composition \citep[see][]{Mendis1977, Whipple1989}, but it was found that the uppermost layers of the nucleus of comet 1P/Halley consist of extremely dark, carbon-rich materials \citep{Keller1987}.

The Rosetta mission has profoundly enriched our view of comets and has confirmed that, although these objects are known to be primarily made of water ice, this compound seems to be rare at the surface of cometary nuclei. As a matter of fact, while water ice was unambiguously detected in the coma of 67P/C-G \citep{Biver2015}, the Visible, InfraRed, and Thermal Imaging Spectrometer (VIRTIS) instrument \citep{Coradini1998} onboard the Rosetta spacecraft has revealed that the surface of 67/C-G is covered by nonvolatile organic materials and that water ice is exposed at the surface only at a few locations, especially in the active regions around the so-called neck of the comet \citep{Capaccioni2015}. Elsewhere, water ice is most likely found under a dehydrated, carbon-rich layer of varying thickness. In line with these findings, at the final landing site of Philae, known as Abydos \citep{Biele2015}, measurements from the MUlti PUrpose Sensors for surface and subsurface Science (MUPUS) instrument package \citep{Spohn2007} have shown that the mechanical and thermal properties of the near-surface layers are consistent with a hard, dust-rich ice substrate covered by a thin dust layer \citep{Spohn2015}. SESAME-PP is one of the few Rosetta instruments that can help to confirm the presence of water ice in the first meters of the subsurface, keeping in mind that water ice may be present in a very porous form. This is suggested by observations from the COmet Nucleus Sounding Experiment by Radio wave Transmission (CONSERT) bistatic radar \citep{Kofman2007}, which found that the average porosity of the smaller lobe of the comet (informally called the head) ranges between 75\% and 85\% \citep{Kofman2015} in agreement with the low density of the nucleus \citep{Sierks2015}.

Permittivity probes derive the electrical properties of a surface material through the coupling of two dipoles located close-to or on the surface. Unlike microwave techniques (e.g., radar), permittivity probes operate at very low frequencies, typically between $10$\,Hz and $10$\,kHz ( Extremely Low Frequency to Very Low Frequency domain). In this frequency range, the electrical signature of a material is especially sensitive to the presence of water ice and its temperature behavior \citep[see][]{Auty1952, Mattei2014}. On November 13 2014, the second day of the First Science Sequence (FSS) phase, SESAME-PP performed four identical sets of measurements on the comet surface at Abydos, the final landing site of Philae. This paper is dedicated to the analysis of the active data that were then collected. The instrument also conducted passive measurements to monitor the plasma waves environment of the nucleus but the analysis of these data lies outside the scope of the present work.

The complex permittivity at low frequency of an extraterrestrial surface was investigated in situ only once before: by the Permittivity, Waves and Altimetry: Huygens Atmospheric Structure Instrument (PWA-HASI) onboard the ESA Huygens lander on the surface of Titan, the largest moon of Saturn \citep{Fulchignoni2005,Grard2006,Hamelin2016}. The same instrument was successfully for permittivity measurements in terrestrial snow of low density \citep{Trautner2003}. In the frame of the PWA-HASI experiment, a numerical method, called the Capacity-Influence Matrix Method was developed to analyze permittivity probe data. Among others, this method has the advantage of taking the operation configuration  into account, which, in the case of Philae at Abydos, was far from nominal. We describe this method and the theory behind surface permittivity probes in Sect. \ref{Section1}. Following a brief description of the instrument, we present the numerical model that was developed for SESAME-PP in Sect. \ref{Section2}. Sect. \ref{Section3} gives an overview of the data that were acquired from the beginning of the Rosetta mission. The FSS data are then analyzed in Sect. \ref{Section4} by comparison with data acquired during the descent toward the nucleus and with numerical simulations that explore different possible attitudes and environments of Philae at its final landing site. Lastly, our results are discussed in Sect. \ref{Section5} in light of other instrument findings.


\section{Theory of surface permittivity probes}\label{Section1}

The SESAME-PP instrument is a Mutual Impedance Probe (MIP) designed to operate at or near the surface, and is based on the quadrupole array technique described below.

\subsection{History and principle of surface MIP}\label{Section11}

MIP have been used on Earth for many decades to measure the subsurface resistivity in a nondestructive way. They were first introduced by \cite{Wenner1916} and consist of four electrodes. In their early version, a DC current was injected between two transmitting electrodes and the potential difference induced by this current was measured between two receiving electrodes in contact with the ground. The ratio of the received voltage potential over the injected current, i.e., the mutual impedance of the quadrupole, yields the conductivity of the subjacent ground down to a depth comparable to the separation between the electrodes (see Sect. \ref{Section23}). Compared to the self-impedance technique, the MIP technique is much less sensitive to the presence of heterogeneities in the vicinity of the electrodes and to the quality of the contact between the electrodes and the medium. Alternatively, the electrodes can be buried at various depths below the surface. 

Later, \cite{Grard1990a,Grard1990b} proposed to use the same technique with AC instead of DC signals in order to measure not only the conductivity, but also the dielectric constant, i.e., the complex permittivity of the ground (see Sect \ref{Section12}). This technique, which had been successfully applied in space plasmas \citep{Storey1969} in the frame of many ionospheric and magnetospheric experiments around the Earth \citep[see][]{Chasseriaux1972, Decreau1982, Decreau1987}, was subsequently validated on Earth \citep{Tabbagh1993}. It was first used at the surface of a planetary body with the PWA analyzer \citep{Grard1995}, a unit of the HASI package \citep{Fulchignoni2002} onboard the ESA Huygens probe that landed at the surface of Titan on January 14, 2005 \citep{Fulchignoni2005,Grard2006,Hamelin2016}. Lastly, a laboratory MIP called HP3-PP \citep{Stiegler2011} had been designed to be part of the ExoMars Humboldt surface station, which was ultimately canceled.

\subsection{The complex permittivity of matter}\label{Section12}

In a macroscopic description, the interaction of harmonic electromagnetic waves with a linear isotropic homogenous medium is completely characterized by three constitutive parameters that are usually frequency dependent: the effective dielectric constant $\epsilon$ (in F/m), the effective electrical conductivity $\sigma$ (in S/m), and the magnetic permeability $\mu$ (in H/m). We assume that the latter parameter is equal to the permeability of vacuum, which is reasonable for natural and icy surfaces that are often nonmagnetic. As a further argument for this hypothesis, the ROsetta MAgnetometer and Plasma monitor (ROMAP) instrument onboard Philae has revealed the nonmagnetic nature of 67P/C-G \citep{Auster2015}. Further, in a harmonic regime and using complex notation, it is common to refer to the relative complex permittivity of the medium (relative to that of vacuum) defined for fields and potentials in $\exp(j\omega t)$ as follows:

\begin{equation}
\epsilon_\mathrm{cplx} = \epsilon_\mathrm{r} - \mathrm{j} \frac{\sigma}{\epsilon_\mathrm{0} \omega}
\label{Eq1}
,\end{equation}

\noindent where $\epsilon_\mathrm{r}$ is the dielectric constant of the medium (i.e., $\epsilon$ normalized by the permittivity of vacuum $\epsilon_\mathrm{0} =8.854\cdot 10^{-12}$\,F/m, $\epsilon_\mathrm{r}=\epsilon/\epsilon_0$) and $\omega$ is the angular frequency (i.e., $\omega=2\pi f$, where $f$ is the working frequency).

From a microscopic point of view, the complex permittivity describes the polarization mechanisms at play in a dielectric material when an electrical field is applied. The main polarization mechanisms, in order of appearance, include: electronic polarization, atomic polarization, orientation (or dipolar) polarization, and space charge (or interfacial) polarization, especially when multiple phase materials or interfaces are involved \citep{Kingery1976}. The respective contribution to the permittivity of these polarization mechanisms depends on the frequency of the applied electric field. The lower the frequency, the more these mechanisms contribute and the higher the dielectric constant. Indeed, at high frequencies, space charge polarization is usually unable to follow changes in the electrical field that occur much faster than the movement of charge carriers. Similarly, at high frequencies, any (permanent or temporary) dipoles present are usually not able to rotate in time with the oscillations of the electric field and orientation polarization thus does not contribute to the total polarizability of matter. In contrast, electronic and atomic polarizations, owing to the displacement of the electronic cloud or of charges in bound atoms to balance the electric field, respectively, have very small response times and, therefore, contribute to the permittivity at all frequencies.
 
The dielectric constant of a material also depends on its temperature because molecular vibrations and polarization are related. At low frequencies, when the temperature decreases, the orientation and space charge polarization mechanisms become less efficient because the dipoles and charge carriers react more slowly to the changes in the electrical field orientation. This results in a decrease of the dielectric constant as illustrated by the case of water ice shown in Sect. \ref{Section13}.

The imaginary part of the complex permittivity depends on the true conductivity that represents the capacity of motion of free charges in matter. It also decreases as the temperatures decreases. We note that while pure dielectric materials have, by definition, a zero true conductivity (i.e., no free charge), polarization mechanisms induce the heat dissipation of electrical energy in matter, which gives rise to an effective conductivity that contributes to $\sigma$. Most natural surfaces are composed of lossy dielectric material. The imaginary part of the complex permittivity varies by multiple orders of magnitude contrary to the real part, which roughly ranges between $1$ (vacuum) and $100$ (water ice at $-20\ ^{\circ}$C at low frequency; see Sect. \ref{Section13}) for natural
materials.

Lastly, following \cite{Grard1990a}, we emphasize that the simultaneous determination of the dielectric constant and conductivity with similar accuracy requires that we operate at a frequency on the order of

\begin{equation}
f_0 = \frac{\sigma}{2 \pi \epsilon_0 \epsilon_\mathrm{r}}
\label{Eq2}
\end{equation}

\noindent for which the real and imaginary parts of the complex permittivity are equal. On Earth, the electrical properties of the ground are dependent on temperature and very dependent on moisture content; they are typically $\epsilon_\mathrm{r}=20$ and $\sigma =10^{-2}$\,S/m for rocks and sediments, which implies an optimal working frequency of $\approx 10$\,MHz. On an icy body at cryogenic temperatures and, in particular, on the surface of a comet, the expected conductivity is much lower ($10^{-8} - 10^{-5}$\,S/m), which leads to a much lower optimal working frequency of $10 - 10^4$\,Hz.

\subsection{The case of water ice}\label{Section13}

The case of water ice is of great interest for the study of comets. In its pure form, this compound has well-known electrical properties that have been investigated by many authors (see \cite{Petrenko1999} for a comprehensive review and, more recently, \cite{Mattei2014}). In the frequency range of the SESAME-PP instrument ($10-10^4$\,Hz), the relative complex permittivity of water ice is well described by the Debye model \citep{Debye1929},

\begin{equation}
\epsilon_\mathrm{cplx} = \epsilon_\mathrm{r\infty} + \frac{\epsilon_\mathrm{rs}- \epsilon_\mathrm{r\infty}}{1+\mathrm{j}\omega \tau}
\label{Eq3}
,\end{equation}
        
\noindent where $\epsilon_\mathrm{r\infty}$ is the relative high-frequency limit permittivity, $\epsilon_\mathrm{rs}$  the static (low-frequency limit) relative permittivity and $\tau$ the relaxation time of water ice in seconds. The value $\tau$ characterizes the delay of establishment of the polarization mechanisms in response to the applied electrical field.

The relative high-frequency limit dielectric constant $\epsilon_\mathrm{r\infty}$ has a slight temperature dependence that can be approximated by a linear function \citep{Gough1972}

\begin{equation}
\epsilon_\mathrm{r\infty}(T) = 3.02+6.41\cdot 10^{-4}T
\label{Eq4}
.\end{equation}

In contrast, the static permittivity $\epsilon_\mathrm{rs}$ is highly dependent on the temperature; it follows an empirical law established by Cole in 1969 \citep{Touloukian1981}, i.e.,

\begin{equation}
\epsilon_\mathrm{rs}(T) = \epsilon_\mathrm{r\infty} + \frac{A_\mathrm{c}}{T-T_\mathrm{c}}
\label{Eq5}
,\end{equation}
        
\noindent where $T$ is the temperature in Kelvin, $T_\mathrm{c}=15$\,K and $A_\mathrm{c}=2.34\cdot 10^4$\,K was determined by fitting Eq. \ref{Eq5} to experimental data for temperatures in the range $200-270$\,K \citep{Johari1978}.

The relaxation time of water ice $\tau$ is also temperature dependent; it increases when temperature decreases following the empirical Arrhenius' law, as determined experimentally over the range of temperature from 200 K to 278 K by \cite{Auty1952} and \cite{Kawada1978} as follows:

\begin{equation}
\tau (T) = A \exp\left(\frac{E}{k\mathrm{_B} T}\right)
\label{Eq6}
,\end{equation}
        
\noindent where $k_\mathrm{B}$ is the Boltzmann constant in eV/K ($k_\mathrm{B}=8.6173324\cdot 10^{-5}$\,eV/K), $E = 0.571$\,eV is the activation energy of water ice, and $A = 5.30\cdot 10^{-16}$\,s is the period of atomic vibrations \citep{Chyba1998}.
        
Separating the real and imaginary parts in Eq. \ref{Eq3} yields        
\begin{equation}
\epsilon_\mathrm{r}(\omega, T) = \epsilon_\mathrm{r\infty} + \frac{\epsilon_\mathrm{rs}(T)- \epsilon_\mathrm{r\infty}}{1+\omega^2 \tau(T)^2}
\label{Eq7}
\end{equation}

\noindent and   

\begin{equation}
\sigma(\omega, T) = \omega^2\tau \epsilon_0 \frac{\epsilon_\mathrm{rs}(T)- \epsilon_\mathrm{r\infty}}{1+\omega^2 \tau(T)^2}
\label{Eq8}
.\end{equation}

\begin{figure}
\centering
\includegraphics[width=8.8cm]{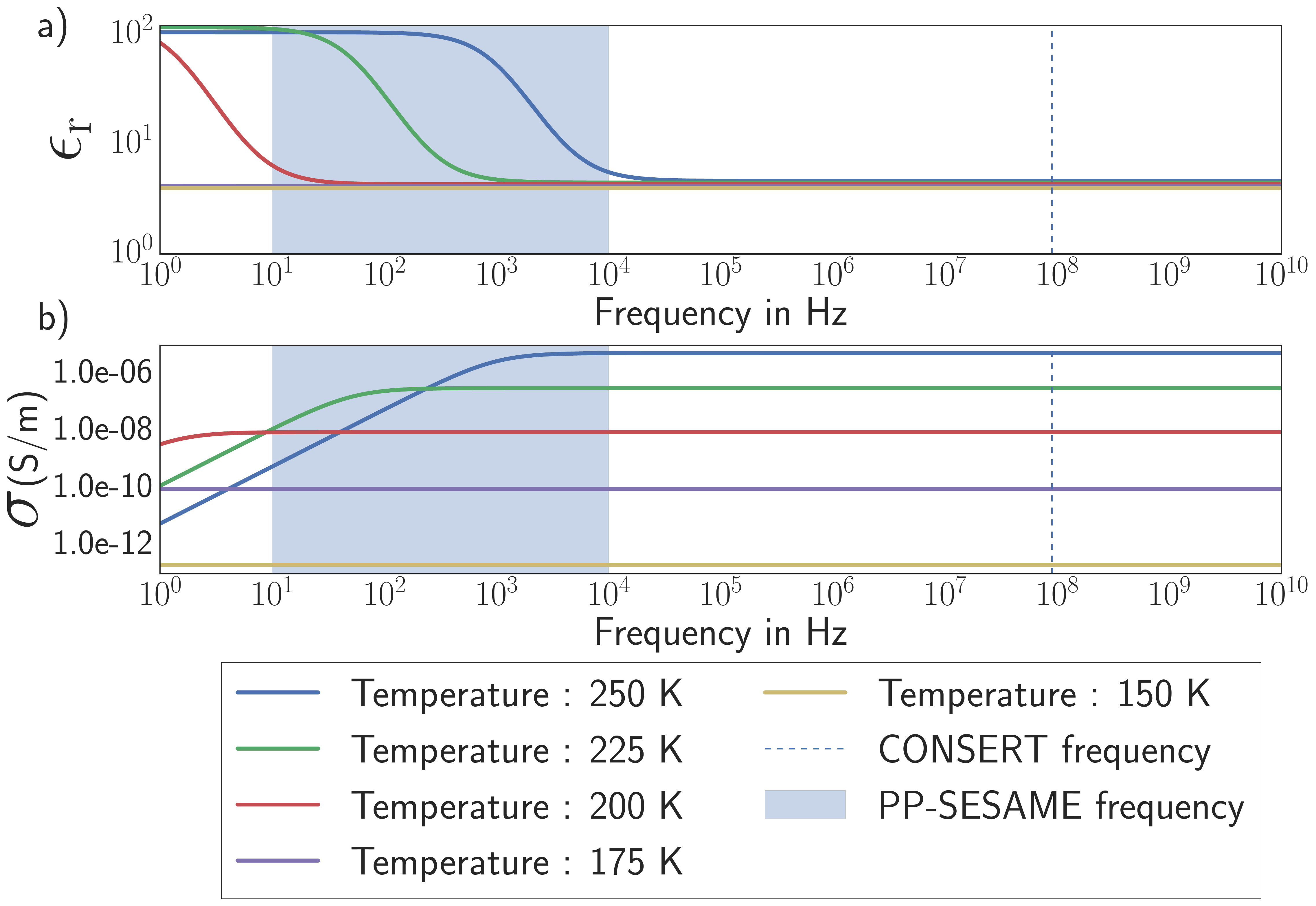}
\caption{Dielectric constant (a) and electrical conductivity (b) of pure water ice as a function of frequency and temperature. The respective operating frequencies of SESAME-PP and CONSERT bistatic radar are indicated.}
\label{Fig1}
\end{figure}

The variations with temperature and frequency of the electrical properties of pure water ice as described by Eq. \ref{Eq7} and Eq. \ref{Eq8} are shown in Fig. \ref{Fig1} (after extrapolation at low temperatures). These equations provide a fair estimate of the dielectric constant and losses of pure water ice.  However, we note that the presence of impurities may significantly affect their validity and, in particular, increase the conductivity.

In the SESAME-PP operating frequency range, the dielectric constant rapidly decreases with temperature, ranging from $\sim 100$ at $250$\,K to $\sim 3.1$ below $175$\,K. This is not the case at higher frequencies and, in particular, at CONSERT operating frequency ($90$\,MHz) for which the relative dielectric constant of water ice can be regarded as constant, equal to about $3.1$. Of importance for the analysis of the data collected over a cold object, such as 67P/C-G, we highlight that below $\sim 175$\,K, or $150$\,K according to the laboratory measurements conducted by \citet{Mattei2014}, the temperature does not affect the relative dielectric constant of water ice anymore, which remains equal to the high-frequency limit value, i.e., $\sim 3.1$. This is due to a very long relaxation time that occurs at cryogenic temperatures. 
 
The value of the water ice dielectric constant at low frequencies ($10$\,Hz to $10$\,kHz) and for a moderately low temperature ($200$\,K to $250$\,K) is especially high (between $10$ and $100$, as shown in  Fig. \ref{Fig1}) compared to other planetary surface materials (most of these have a relative dielectric constant lower than $10$) and this is the reason why surface permittivity probes have been used for the detection and estimation of the subsurface water ice content. High-frequency measurement devices such as radars are better suited for the detection of liquid water, which has a high dielectric constant (80) in the microwave domain due to reorientation of the molecular H2O dipoles. It therefore follows that these two techniques are complementary.

The conductivity of water ice strongly varies with temperature at all frequencies. It decreases when the temperature decreases and progressively loses its frequency dependence. We also note that the conductivity increases with the degree of impurity of the ice.

In this regard, we highlight that several mixing laws have been proposed for the electrical properties of mixed materials. In particular, the Maxwell-Garnett mixing formula \citep{Sihvola1999} states that the effective dielectric constant $\epsilon_\mathrm{eff}$ of a three-phase mixture, consisting of vacuum and ice spherical inclusions in a dusty matrix, can be obtained from

\begin{equation}
\frac{\epsilon_\mathrm{eff}- \epsilon_\mathrm{dust}}{\epsilon_\mathrm{eff}+ 2\epsilon_\mathrm{dust}}= f_\mathrm{ice} \frac{\epsilon_\mathrm{ice}- \epsilon_\mathrm{dust}}{\epsilon_\mathrm{ice}+ 2\epsilon_\mathrm{dust}} + f_\mathrm{dust} \frac{1- \epsilon_\mathrm{dust}}{1+ 2\epsilon_\mathrm{dust}}
\label{Eq9}
,\end{equation}
        
\noindent where $\epsilon_\mathrm{ice}$ is the dielectric constant of the ice, $\epsilon_\mathrm{dust}$ the dielectric constant of the dust, $f_\mathrm{ice}$ the volumetric fraction of ice, and $f_\mathrm{dust}$ the volumetric fraction of dust. 

\subsection{Mutual impedance for a quadrupole above a half-space with a given complex permittivity}\label{Section14}

We summarize the theory of the quadrupolar array to show the principle of the derivation of the complex permittivity $\epsilon_\mathrm{cplx}$ of a planetary surface. This approach was first proposed by \citet{Grard1990a,Grard1990b} and \citet{Grard1991}. It assumes quasi-static approximation, as the wavelength of operation is much larger than the distance between the electrodes, and neglects magnetic induction.

In vacuum, the potential $V$ at a distance $r$ from a point charge $Q$ is

\begin{equation}
V = \frac{Q}{4\pi \epsilon_0 r}
\label{Eq10}
.\end{equation}
        
When this charge is at a height $h$ above an interface separating vacuum from a half-space of relative complex permittivity $\epsilon_\mathrm{cplx}$, the potential distribution can be determined with the image charge theory in which we evaluate the effect of the interface by an image charge located at a distance $h$ under the interface. The charge of the image is equal to \citep{Griffiths2012}
        
\begin{equation}
Q^\prime = -\frac{\epsilon_\mathrm{cplx} - 1}{\epsilon_\mathrm{cplx} + 1} Q = -\alpha_\mathrm{m} Q
\label{Eq11}
.\end{equation}
        
\noindent The potential of a point located above the interface is then
        
\begin{equation}
V = \frac{Q}{4\pi \epsilon_0} \left(\frac{1}{r} - \frac{\alpha_\mathrm{m}}{r^\prime}\right)
\label{Eq12}
,\end{equation}
        
\noindent where $r^\prime$ is the distance between the point and image of the charge.
        
\begin{figure}
\centering
\includegraphics[width=8.8cm]{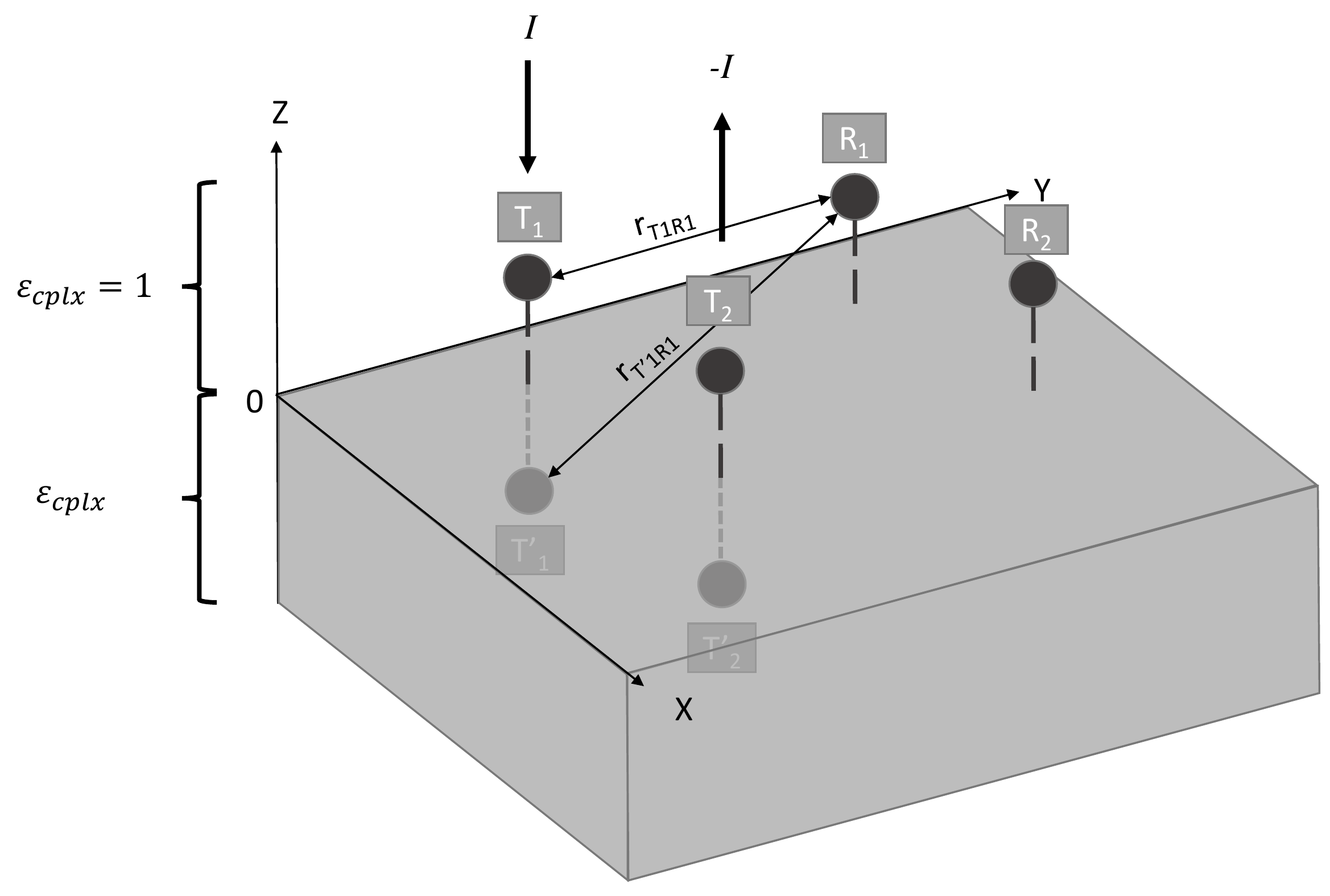}
\caption{Quadrupolar array above an interface separating a half-space with a relative complex permittivity $\epsilon_\mathrm{cplx}$ and vacuum. {$T_1$ and $T_2$} are the transmitting electrodes, while $R_1$ and $R_2$ are the receiving electrodes. $T^{\prime}_1$ and $T^{\prime}_2$ are the images of the transmitting electrodes by the interface. The parameter $r_{T_nR_m}$ is the distance between the $T_n$ and $R_m$ electrodes. The parameter $I$ is the current flowing through the transmitting electrodes.}
\label{Fig2}
\end{figure}

We now consider a system of four pinpoint electrodes located above an interface separating a half-space with a uniform relative complex permittivity $\epsilon_\mathrm{cplx}$ and vacuum as illustrated by Fig. \ref{Fig2}. A sinusoidal current $I$ of angular frequency $\omega$ is fed into the two transmitting electrodes ($T_1$ and $T_2$). In the harmonic regime, $I=I_0 \exp(\mathrm{j}\omega t)$ and the charge $Q= \int Idt$ applied on a transmitting electrode is then
        
\begin{equation}
Q = \frac{I}{\mathrm{j}\omega}=  \frac{I\exp(-\mathrm{j}\pi/2)}{\omega}
\label{Eq13}
.\end{equation}
        
Using Eq. \ref{Eq12} and the theorem of superposition, the potentials induced on the receiving electrodes ($R_1$ and $R_2$) can be written as
        
\begin{equation}
V = \frac{Q}{4\pi \epsilon_0 } \left[\left(\frac{1}{r_{T_2R_1}} - \frac{1}{r_{T_1R_1}}) - \alpha_\mathrm{m} (\frac{1}{r_{T^{\prime}_2R_1}} - \frac{1}{r_{T^{\prime}_1R_1}}\right)\right]
\label{Eq14}
,\end{equation}
\begin{equation}
V = \frac{Q}{4\pi \epsilon_0 } \left[\left(\frac{1}{r_{T_2R_2}} - \frac{1}{r_{T_1R_2}}) - \alpha_\mathrm{m} (\frac{1}{r_{T^{\prime}_2R_2}} - \frac{1}{r_{T^{\prime}_1R_2}}\right)\right]
\label{Eq15}
,\end{equation}
        
\noindent where $r_{T_m R_n}$is the distance between the transmitting $T_m$ and the receiving $R_n$ electrodes and $r_{T^{\prime}_m R_n}$ is the distance between the image $T^{\prime}_m$ (image of the transmitting electrode $T_m$) and the receiving electrode $R_n$.

The mutual impedance of the quadrupole can therefore be written as

\begin{equation}
\begin{array}{c}
\displaystyle{Z_m = \frac{\Delta V}{I} = \frac{V_{R_2} - V_{R_1}}{I}}\\
\displaystyle{=  \frac{1}{4\pi \epsilon_0 \omega } \biggl[\left(\frac{1}{r_{T_1R_1}} +\frac{1}{r_{T_2R_2}}  - \frac{1}{r_{T_1R_2}}  -  \frac{1}{r_{T_2R_1}}\right)}\\
\displaystyle{- \alpha_\mathrm{m} \left( \frac{1}{r_{T^{\prime}_1R_1}} +\frac{1}{r_{T^{\prime}_2R_2}}  - \frac{1}{r_{T^{\prime}_1R_2}}  -  \frac{1}{r_{T^{\prime}_2R_1}} \right) \biggr]}
\label{Eq16}
\end{array}
,\end{equation}
        
\noindent where $\Delta V$ is the potential difference between the two receiving electrodes.   
        
Normalizing $Z_\mathrm{m}$ by $Z_0$, the mutual impedance in vacuum (corresponding to $\alpha_\mathrm{m}=0$), we further obtain

\begin{equation}
\frac{Z_\mathrm{m}}{Z_0} = 1-\delta \alpha_\mathrm{m}
\label{Eq17}
,\end{equation}
        
\noindent with $\delta$, the quadrupole geometrical factor, defined as
        
\begin{equation}
\begin{array}{c}
\displaystyle{
\delta = (\frac{1}{r_{T_1R_1}} +\frac{1}{r_{T^{\prime}_2R_2}}  - \frac{1}{r_{T^{\prime}_1R_2}}  -  \frac{1}{r_{T^{\prime}_2R_1}})}\\
\displaystyle{/(\frac{1}{r_{T_1R_1}} +\frac{1}{r_{T_2R_2}}  - \frac{1}{r_{T_1R_2}}  -  \frac{1}{r_{T_2R_1}})}
\label{Eq18}
\end{array}
,\end{equation}
        
The complex permittivity of the lower half-space can then be derived from measurements of both the mutual impedance in vacuum and above the half-space using the following equation:
        
\begin{equation}
\epsilon_\mathrm{cplx} = \frac{Z_0 (\delta + 1) - Z_\mathrm{m}}{Z_0 (\delta - 1) - Z_\mathrm{m}}
\label{Eq19}
.\end{equation}
        
However, Eq. \ref{Eq19} has limitations: it applies well to isolated idealized pinpoint electrodes, but does not account for the effect of the close environment of the MIP (for instance, in the case of Philae, the presence of the lander body) nor for the electronic circuit that links the electrodes together and for the shape of the electrodes. For a more realistic approach, we adapted a method called the Capacitance-Influence Matrix Method, which was successfully applied to the analysis of the data collected by PWA-HASI/Huygens at the surface of Titan \citep{Hamelin2016}.

\subsection{The Capacitance-Influence Matrix Method}\label{Section15}

The Capacitance-Influence Matrix Method is based on the lumped element model, which consists in representing the electrical characteristics of the MIP environment as a network of fictive lumped elements. These conducting elements could, additionally, be linked together by an electronic circuit. This approach allows us to establish and use for prediction a set of linear equations that describes the whole system.

We consider the case of N disconnected conductors in a dielectric medium. By superposition, the charge $Q_k$ on the $k^{th}$ conductor due to the $N$ charged conductors in the system is given by
        
\begin{equation}
Q_k = \sum_{n=1}^{N}{K^\mathrm{m}_{kn}V_n}
\label{Eq20}
,\end{equation}

\noindent where $K^\mathrm{m}_{kn}$ with $n=0,1,2,\ldots, N$ are the coefficient of the medium capacitance-influence matrix $\vec{[K^\mathrm{m}]}$ and $V_n$ is the potential on the $n^{th}$ conductor. This results in the matricial equation

\begin{equation}
\vec{Q} = \vec{[K^\mathrm{m}]}\vec{V}
\label{Eq21}
\end{equation}

\noindent and using Eq. \ref{Eq13}

\begin{equation}
\vec{I} = \mathrm{j} \omega \vec{[K^\mathrm{m}]}\vec{V}
\label{Eq22}%
,\end{equation}

\noindent where $\vec{Q}=[{Q_1,Q_2,\ldots,Q_N}]$ is the vector of the charges on the discrete conductors, $\vec{V}=[{V_1,V_2,V_3,\ldots,V_N}]$ the vector of the potentials of the discrete conductors,  and $\vec{I}=[{I_1,I_2,I_3,\ldots,I_N}]$  the vector of currents injected into the medium by the electronic circuit that is represented by its electronic admittance matrix $\mathrm{j}\omega \vec{[K^\mathrm{e}]}$. The system, composed of the electronic circuit and the medium, is represented by the equations

\begin{equation}
\vec{I} = j \omega \vec{[K]}\vec{V}
\label{Eq23}
\end{equation}

\noindent
or 
\begin{equation}
\vec{V} = \vec{[K]}^{-1} \frac{\vec{I}}{\mathrm{j}\omega}
\label{Eq24}
\end{equation}

\noindent
with
\begin{equation}
\vec{[K]}=\vec{[K^\mathrm{e}]}+\vec{[K^\mathrm{m}]}
\label{Eq25}  
.\end{equation} 

The matrix $\vec{[K]}$ is the capacitance-influence matrix of the multiconductor system, and $\vec{[K^\mathrm{m}]}$ can be obtained by modeling. In practice, we build a numerical geometry model of the instrument and its conductive environment, which includes the planetary dielectric surface, and we use the software COMSOL Multiphysics® to solve the Laplace equations (see Sect. \ref{Section222}). \ The matrix $\vec{[K^\mathrm{m}]}$ varies with the configuration of operation (location/attitude of the electrodes with respect to the surface) and with the complex permittivity of the surface material. The matrix $\vec{[K^\mathrm{e}]}$ is obtained from the electronic circuit analytical model (see Sect. \ref{Section223}).

The Capacitance-Influence Matrix Method consists in using Eq. \ref{Eq24}, which corresponds to a set of $N$ equations, with some additional constraints on the vectors $\vec{I}$ and $\vec{V}$ (see Sect. \ref{Section224}) to predict the potentials on the receiving electrodes for a variety of planetary surface electrical properties. These predictions can then be compared to the data to find the complex permittivity that best reproduces the observations; namely, the measured received potentials and/or their difference, where the injected current is measured by SESAME-PP. In this approach, the main source of uncertainty is $\vec{[K^\mathrm{e}],}$ which is subject to change with time because of aging, for instance. In addition, the derivation of $\vec{[K^\mathrm{m}]}$ requires good knowledge of the configuration of operation of the MIP.

\section{The SESAME-PP/Philae instrument}\label{Section2}

\subsection{Description}\label{Section21}

\begin{figure}
\centering
\includegraphics[width=8.8cm]{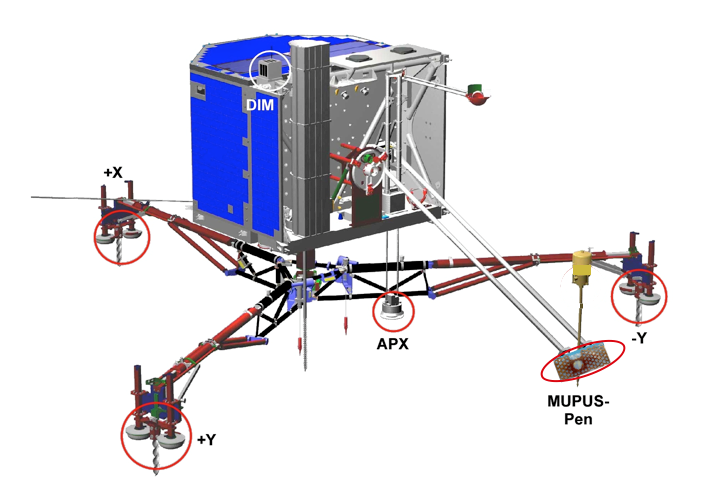}
\caption{ Rosetta lander Philae with deployed landing gear and appendages showing the locations of the SESAME-PP sensors (red circles). Three electrodes are located on the feet of the lander; each of these electrodes is composed of two interconnected soles. The two other electrodes are colocated with the MUPUS-PEN and APXS instruments. Copyright ESA/ATG medialab.}
\label{Fig3}
\end{figure}

The SESAME-PP instrument is part of the SESAME package \citep{Seidensticker2007}; it shares most of its electronics and has a common software with the Dust Impact Monitor (DIM) and Cometary Acoustic Surface Sounding Experiment (CASSE) experiments \citep{Seidensticker2007,Kruger2015a}. SESAME-PP is composed of five electrodes, three transmitting and two receiving (see Fig. \ref{Fig3}). The two receiving electrodes are located on two of the feet of the Philae lander; hereafter called +Y and $-$Y electrodes. One of the transmitting electrodes is located on the third foot of the lander hereafter referred to as +X; the two other are collocated with the Penetrator of MUPUS (MUPUS-PEN) instrument \citep{Spohn2007} and AlPha X-ray Spectrometer (APXS) sensor \citep{Klingelhofer2007}. The APXS electrode and deployment device are mounted in an opening in the floor of the balcony of the Philae body, while the MUPUS-PEN electrode is designed to be deployed up to $1$\,m away from the back of the lander body.

In the active mode of operation of SESAME-PP, two transmitting electrodes are selected to inject a current into the environment and the induced potential difference between the two receiving electrodes is measured. Additionally, a second signal can be acquired: this signal can either be the current that flows through one of the two transmitting electrodes or the potential sensed by one of the receiving electrodes. By default, the data consist of the amplitude and phase of the potential difference or current; these two quantities are obtained after processing  the measured signals onboard. However, time series can also be acquired; the data are then processed on Earth at the cost of a larger data volume. 

The transmitting dipole consists either of +X and MUPUS-PEN, +X and APXS, or MUPUS-PEN and APXS. Though, in principle, only four electrodes are required, the possibility of selecting among three transmitting dipoles and  varying the geometry of the quadrupole allows us to probe different volumes of material and possibly to detect heterogeneities in the near subsurface. A reduced geometry mode that makes use of the three foot electrodes only (+X as the transmitter and $-$Y and +Y as the receivers) was also anticipated for in-flight calibration and for the first measurements after landing before the deployments of APXS and MUPUS. In this mode of operation, the body of Philae acts as the second transmitting electrode, which is not the optimal situation. Unfortunately, because of the shortness the Philae mission, the measurements performed on the surface of 67P/C-G were only acquired in this mode of operation (see Sect. \ref{Section33}).

The SESAME-PP instrument also includes a passive operation mode that records the potential difference between the two receiving electrodes with a sampling frequency of $40$\,kHz, without any active transmitting electrode. The main objective of this mode is to measure the electric field of the plasma waves generated by the interaction of the solar wind with the charged dust and ionized gases that surround the nucleus, and thus to monitor the activity of the comet. The analysis of the passive measurements collected by SESAME-PP during the descent and on the surface of the nucleus lies outside the scope of this paper.
                        
The block diagram of SESAME-PP is shown in Annex \ref{Fig4}. Upon command, a sinus wave is generated by a digital-to-analog converter; the signal is then smoothed and amplified in two channels in phase opposition. A switch-box then connects the two signals to the selected transmitting electrodes (+X, MUPUS-PEN, APXS, or none). On the receiving side, the signals delivered by the preamplifiers built in each foot are filtered and fed into a differential amplifier. The signals are then selected, sent via a multiplexer to the analog-to-digital converter and processed by the onboard computer.

The SESAME-PP instrument offers three important features. First, its maximum power requirement is $1767$\,mW, each measurement lasts $6$\,s and requires $3$\,mWh of energy (including the power needed by the SESAME computer). Secondly, the required data volume is small: at most, $2816$\,bytes per measurement for the time series. Thirdly, the total mass of the instrument does not exceed $170$\,g and, thus, easily meets space mission requirements.

Lastly, we emphasize that SESAME-PP has the advantage, contrary to PWA-HASI/Huygens, to measure the current injected into the medium, which yields additional information about the self-impedance of the transmitting electrodes. This is an important feature that helps to avoid errors associated with the possible presence of heterogeneities around the transmitting electrodes.  

\subsection{A numerical model for SESAME-PP/Philae}\label{Section22}

\subsubsection{SESAME-PP lumped element model}\label{Section221}

\begin{figure}
\centering
\includegraphics[width=8.8cm]{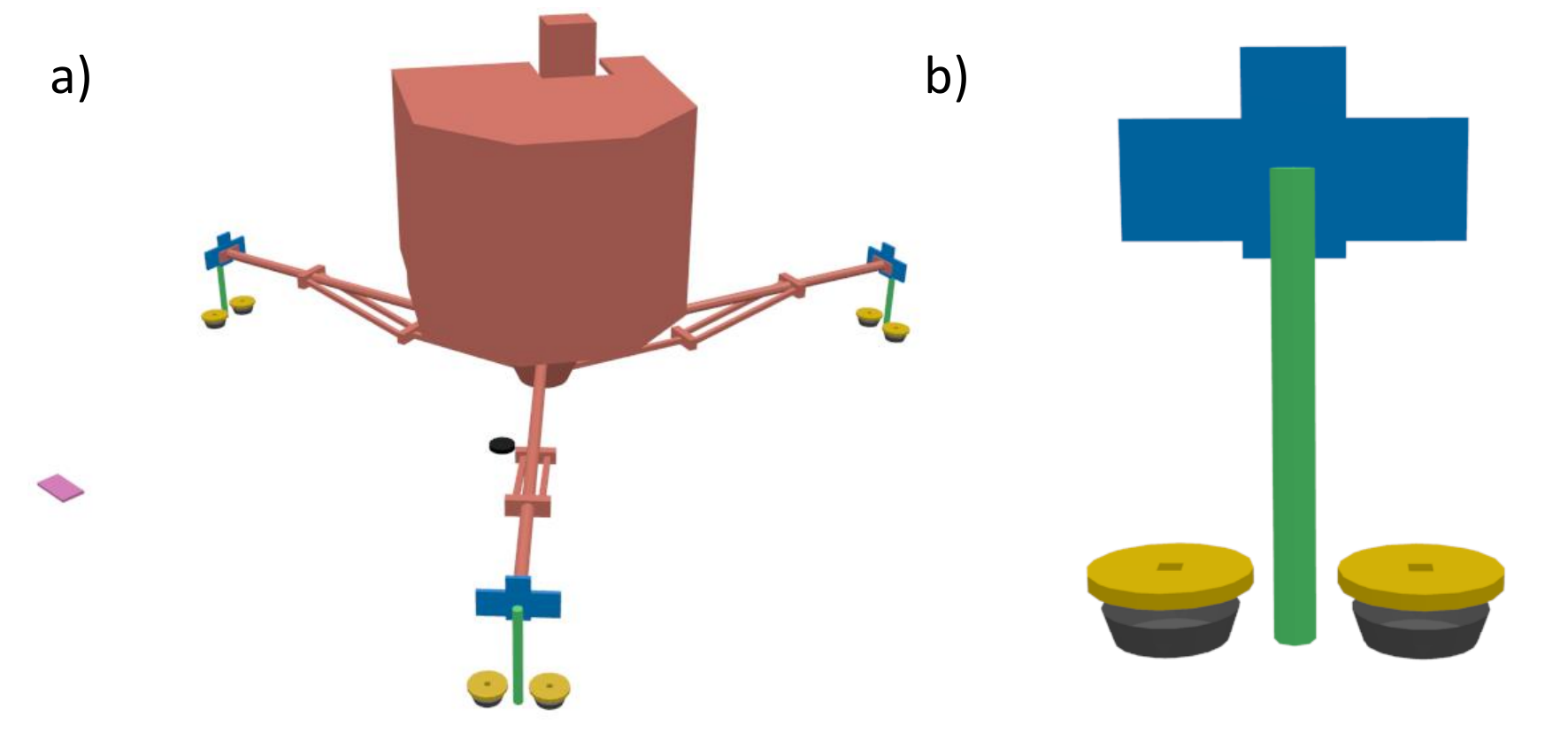}
\caption{(a) Numerical geometry model of the SESAME-PP instrument and its conducting environment. The red shape is the body of the lander combined with the landing gear (element 13). The body is rotated by $11.2\ ^{\circ}$ with respect to the landing gear. The blue shapes are the conductive plates linking the legs of the lander to the feet (elements 1, 5, and 9). The green shapes are the ice screws on each foot (elements 2, 6, and 10). The yellow shapes are the guards of the electrodes (elements 3, 7, and 11). The dark gray shapes represent the soles of the electrodes (elements 4, 8, and 12). The pink shape is the MUPUS-PEN electrode and guard (elements 16 and 17). Lastly, the black shape is the APXS electrode (elements 14 and 15), once deployed under the balcony. Not represented on this numerical model are the harpoons located under the body of the lander (elements 18 and 19). (b) Zoom on one of the feet of the lander.}
\label{Fig5}
\end{figure}

In order to apply the Capacitance-Influence Matrix Method, we discretize the conductive environment of SESAME-PP into a set of $19$ discrete conducting elements. The numerical geometry model is illustrated in Fig. \ref{Fig5} with the $19$ elements shown in Fig. \ref{Fig5}a and listed in Table \ref{table:1}. The Philae body (element number 13) is the largest of these elements. A close-up of the numerical model constructed for the feet of the lander is also shown (Fig. \ref{Fig5}b). The transmitting electrodes are element numbers 3--4 (+X electrode), 14--15 (MUPUS-PEN electrode), and 16--17 (APXS electrode); the receiving electrodes are element numbers 7--8 ($-$Y electrode) and 11--12 (+Y electrode).

For the sake of simplicity and to limit computation time, the geometry of the lander was simplified. Simplifications include:
\begin{enumerate}[i)]
\item The part linking the body to the legs is approximated by conical section.
\item The interface between the legs and feet is replaced by a parallel plate capacitor of equivalent capacitance.
\item The geometrical model of the feet omits the presence of the nonconductive elements holding the soles and the shape of the soles is approximated by the section of a cone.
\item The screws are approximated by cylinders with appropriate dimensions.
\item The CASSE sensors are not included in the soles.
\item The MUPUS-PEN electrode is represented by a block with appropriate dimensions.
\item   The APXS electrode is represented by a cylinder with appropriate dimensions.
\end{enumerate}

The limitations of these simplifying assumptions were tested against simulations using more sophisticated geometries and are proven to have very little impact on the results. The model also offers the possibility of rotating the body of the lander with regard to the landing gear to simulate the attitude of Philae during the descent and on the surface.

\begin{table}
\caption{Conducting elements of SESAME-PP and its conducting environment}             
\label{table:1}      
\centering          
\begin{tabular}{l l l l}     
\hline       
Element number & Element\\
\hline                    
1 & +X foot plate    \\
2 & +X screw \\
3 & +X guards \\
4 & +X soles \\
5 & $-$Y foot plate \\
6 & $-$Y screw  \\
7 & $-$Y guards \\
8 & $-$Y soles   \\
9 & +Y foot plate   \\
10 & +Y screw   \\
11 & +Y guards \\
12 & +Y soles   \\
13 & Body of the lander \& Landing Gear    \\
14 & Guard APXS \\
15 & Electrode APXS  \\
16 & Guard MUPUS-PEN  \\
17 & Electrode MUPUS-PEN    \\
18 & Harpoon 1 \\
19 & Harpoon 2 \\
\hline                  
\end{tabular}
\end{table}

\subsubsection{Derivation of medium matrix $\vec{[K^m]}$}\label{Section222}

\begin{figure}
\centering
\includegraphics[width=8.8cm]{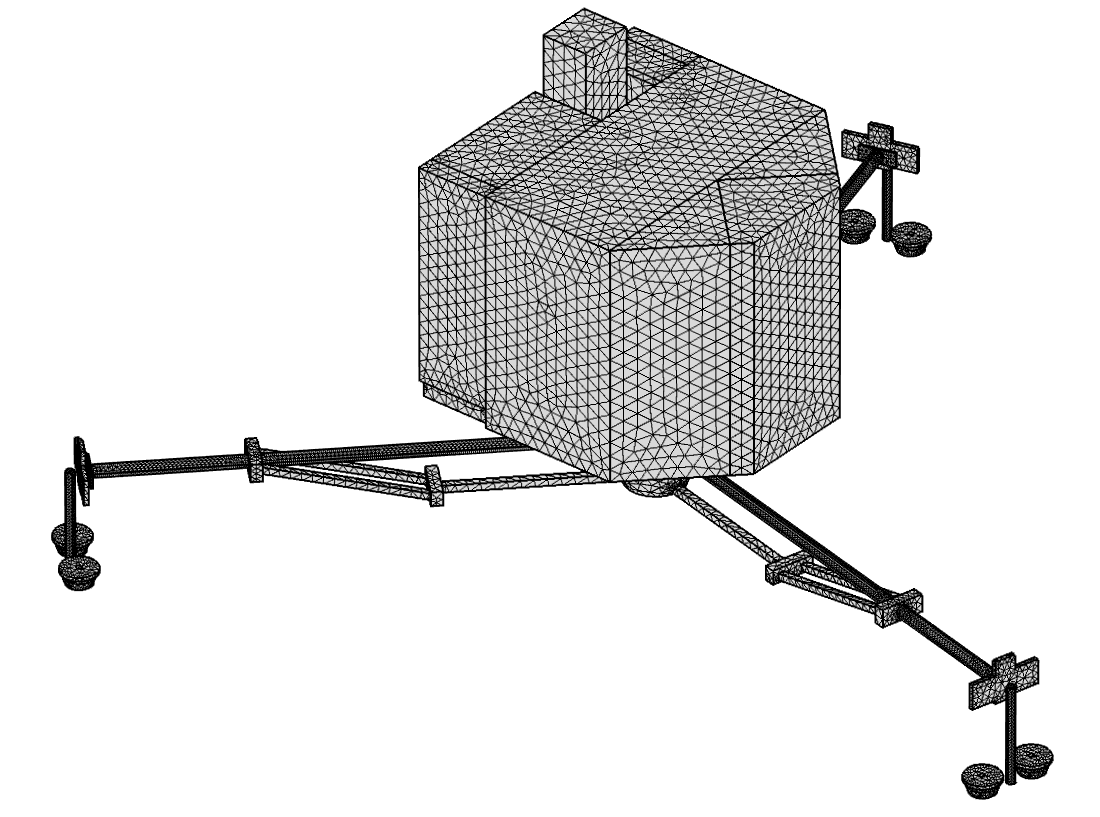}
\caption{Meshed model of the Philae lander}
\label{Fig6}
\end{figure}

In order to derive $\vec{[K^\mathrm{m}]}$, we import a numerical model of the lander (shown in Fig 5) and its attitude with respect to the environment into COMSOL Multiphysics© (https://www.comsol.com). The ground is characterized by its complex permittivity and morphology (the simplest is a plane, but more complex surfaces can be modeled; see Sect. \ref{Section431}). The whole lander with its conducting elements and the environment are then meshed (Fig. \ref{Fig6}) and Dirichlet boundary conditions are set: zero potential at infinity and fixed potentials on the elements of the model ($0$ or $1$\,V). The code then solves the Laplace equation using the finite element method. For an in-depth description of the resolution method, see \citet{Durand1966}.

The code cycles through the $19$ elements. A Dirichlet boundary conditions potential of $1$\,V is applied to the active conductors, while those of the others are set to $0$\,V. COMSOL then calculates the charges $Q$ carried by each element for each cycle and, hence, the matrix $[\vec{K}^\mathrm{m}]$ (Eqs. 20 and 21). The code can be run for a variety of environment models, lander attitudes, and electrode positions on the surface. The complex permittivity of the subsurface is derived by comparison with the simulation outputs with an accuracy that is strongly influenced by our knowledge of the attitude of the lander with respect to the surface. This numerical approach was validated against the results obtained in situations that can be solved analytically   

\subsubsection{Derivation of the electronic matrix  $\vec{[K^\mathrm{e}]}$}\label{Section223}

The conducting elements are also linked by the electronic circuit. The electronic matrix of SESAME-PP is given in Table \ref{table:table2}. The components of this matrix are derived from a combination of

\begin{enumerate}[i)]
\item calibration measurements performed on the flight model prior to launch, 
\item measurements performed with a ground model built at LATMOS (France), and
\item electronic models.
\end{enumerate}

The electronics of the instrument might also have aged during the ten-year Rosetta cruise. For safety, a calibration cycle was planned during the descent of Philae toward the nucleus in a near-vacuum environment. Unfortunately, these measurements were strongly disturbed by interferences (Sect. \ref{Section32}).

\subsubsection{Constraints}\label{Section224}

Once $\vec{[K^\mathrm{m}]}$ and $\vec{[K^\mathrm{e}]}$ are determined, the received potentials and injected current for a given environment are derived from Eq. \ref{Eq24} ($19$ linear equations with $19$ unknowns), taking  the $19$ constraints that apply to the potentials and currents into
account. These constraints with their rational are recapitulated in Table \ref{table:3}.

\begin{table*}
\centering
\caption{Constraints applied to the potentials and currents to solve Eq. \ref{Eq24}}
\label{table:3}
\begin{tabular}{l l l}
\hline
Variable & Constraint & Comments \\ \hline
\begin{tabular}[c]{@{}l@{}}$I_1,I_2,I_3,(I_4),I_5,I_6,I_7,I_8,I_9,I_{10}, I_{11},$ \\$ I_{12}, (I_{13}),I_{14},(I_{15}),I_{16},(I_{17}),I_{18},I_{19}$\end{tabular} & $0$ A & \begin{tabular}[c]{@{}l@{}}The current injected in a passive conductor is zero. \\ The currents in transmitting electrodes (between \\ brackets) are set to 0 only when not used.\end{tabular} \\ \hline
& & \\
$\sum_{k=1}^{19} I_k=0$ & $0$ A & Kirchhoff’s law \\
& &\\ \hline
& & \\
\begin{tabular}[c]{@{}l@{}}$V_3-V_{13}$ or $V_3-V_{15}$ or \\ $V_3-V_{17}$ or $ V_{15}-V_{17}$\end{tabular} & Amplitude of the transmitted signal & \begin{tabular}[c]{@{}l@{}}The potential difference between the two \\ transmitting electrodes is set by telecommand.\end{tabular} \\ 
& & \\ \hline
$V_{11}-V_{13}$ & $(V_{8}-V_{13}) \alpha_{11}$ & \begin{tabular}[c]{@{}l@{}}The potentials of the guards of the receiving \\electrodes  are equal to the potential of the soles\\ multiplied by the transfer factors $\alpha_7$ and $\alpha_{11}$.\end{tabular} \\ \cline{1-2}
$V_7-V_{13}$ & $(V_8-V_{13}) \alpha_{7}$ &  \begin{tabular}[c]{@{}l@{}}The parameters $\alpha_7$ and $\alpha_{11}$ depend on \\temperature and frequency and have been \\measured.\end{tabular} \\ \cline{1-3}
\end{tabular}
\end{table*}

\subsection{Assessment of SESAME-PP performances: Sounding depth and apparent permittivity}\label{Section23}

The performances of SESAME-PP, for example, the equivalent sounding depth, can be assessed with the Capacitance-Influence Matrix Method. For that purpose, we first consider that the lander with the SESAME-PP instrument lies on the surface of a medium with a dielectric constant of $2.30$ (a value close to the inferred lower limit of the dielectric constant of 67P/G-C’s nucleus; see Sect. \ref{Section432}) and we then assume a perfect reflector at different depths below the surface for different positions of the MUPUS-PEN transmitting electrode. 

For each position of the MUPUS-PEN, we estimate the mutual impedance as a function of the reflector depth and consider that the sounding depth corresponds to the depth at which the reflector significantly influences the measurements (i.e., when the difference between the mutual impedance with and without reflector exceeds the error on the impedance measurement, namely $\sim 10\%$).

\begin{figure}
\centering
\includegraphics[width=8.8cm]{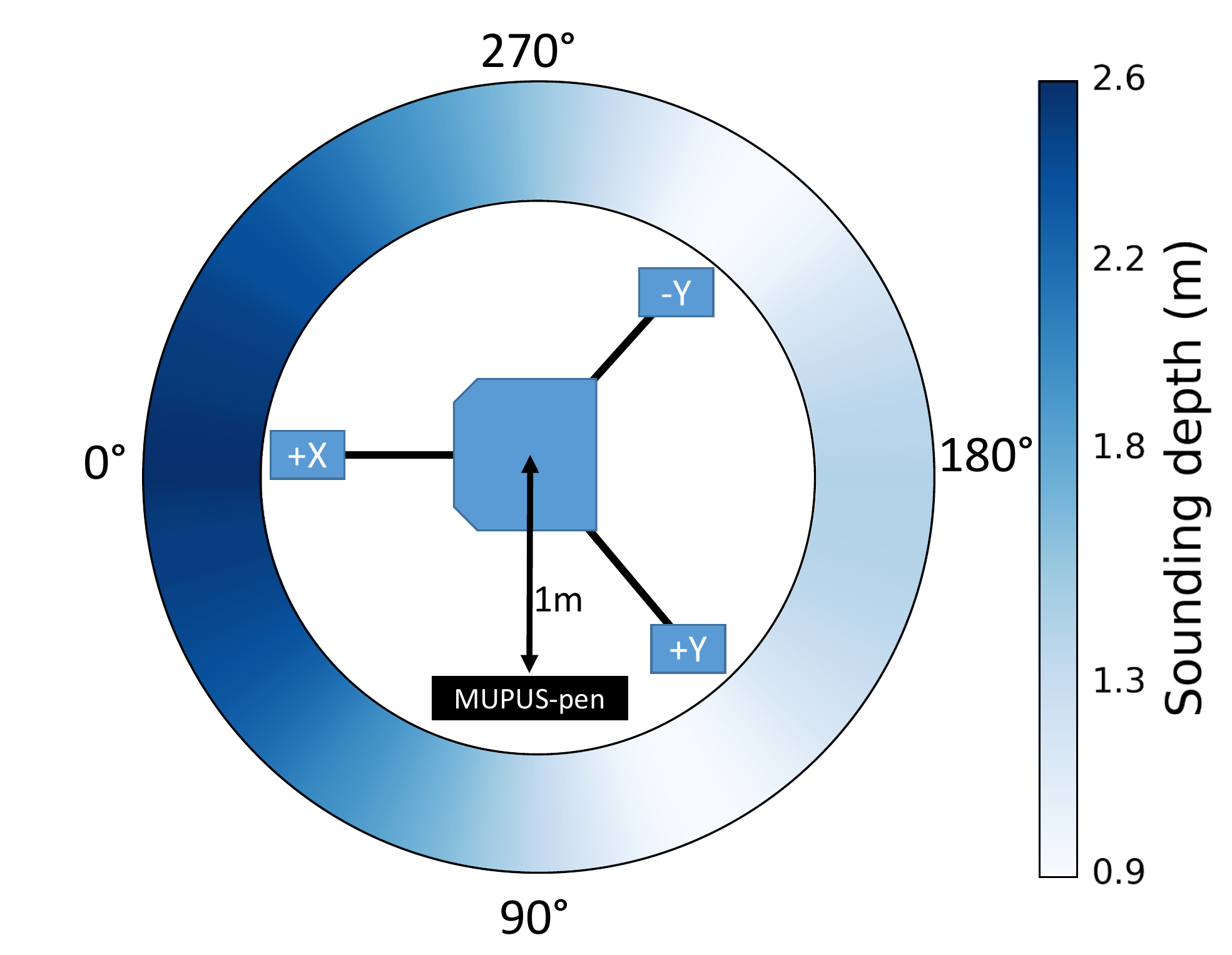}
\caption{Theoretical sounding depth of SESAME-PP as a function of the MUPUS-PEN position (in degrees with respect to the +X leg direction) when the transmitting electrodes are +X and MUPUS-PEN. The distance of $1$\,m indicated for the MUPUS-PEN electrode is the maximum theoretical deployment distance. At Abydos the final deployment distance was $\approx 60$\,cm \citep{Spohn2015}.}
\label{Fig7}
\end{figure}

Fig. \ref{Fig7} shows that the SESAME-PP sounding depth varies with the position of the MUPUS-PEN transmitting electrode at between $0.9$ and $2.6$\,m. By using SESAME-PP with different configurations of operation, different depths below the surface (down to $2.6$\,m) can be sounded and the instrument can therefore detect possible layers in the subsurface. When SESAME-PP is operated with only three foot electrodes, the theoretical sounding depth is found to be $1.3$\,m, while it is reduced to $1.0$\,m with +X and APXS as transmitting electrodes.

\begin{figure}
\centering
\includegraphics[width=8.8cm]{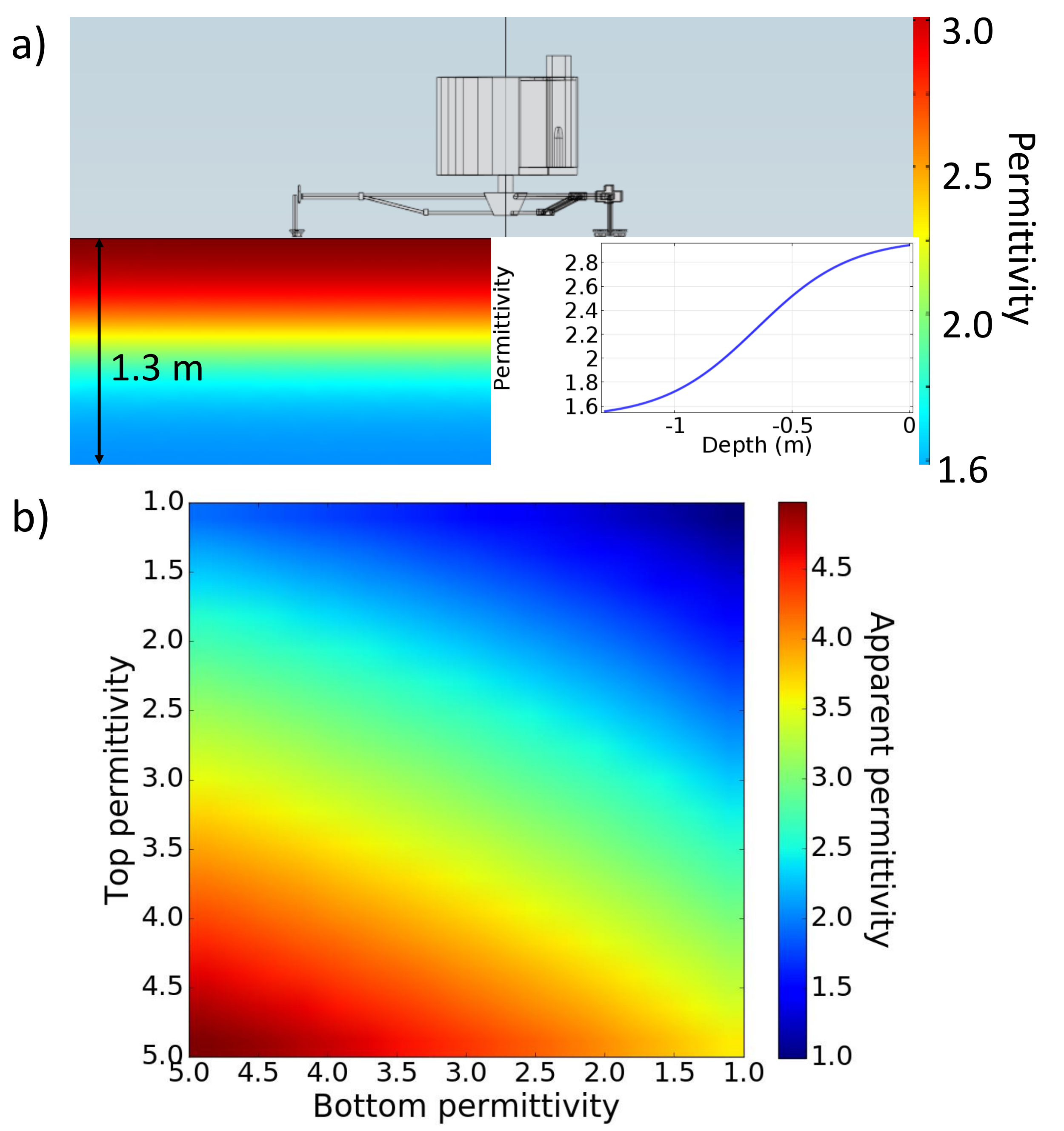}
\caption{(a) SESAME-PP/Philae over a pure dielectric surface exhibiting a permittivity varying as a sigmoid function from $1.6$ at $1.3$\,m below the surface up to $3$ at the surface). (b) Apparent dielectric constant sensed by SESAME-PP as a function of the permittivities on the surface (top permittivity) and at $1.3$\,m below the surface (bottom permittivity).}
\label{Fig8}
\end{figure}

The ability of SESAME-PP to sound varying volumes of the subsurface is a key to the detection of buried heterogeneities or a permittivity gradient below the surface. More specifically, SESAME-PP gives access to an apparent permittivity that is an average of the complex permittivity of the subsurface. As an example, the apparent dielectric constant inferred from the SESAME-PP observations in the case of a pure dielectric subsurface displaying a sigmoid variation of the permittivity (characterized by the dielectric constant at the surface and $1.3$\,m below the surface) is shown in Fig. \ref{Fig8} for the three foot configuration. If an apparent permittivity of $3$ is derived from SESAME-PP observations, only multiconfiguration measurements can help discriminate between a homogeneous subsurface of permittivity $3$, an increasing permittivity gradient, or a decreasing permittivity gradient.
           
\section{SESAME-PP/Philae measurements during the Rosetta mission}\label{Section3}

\subsection{Cruise}\label{Section31}

The SESAME-PP instrument was activated on several occasions (a dozen times) during the cruise phase of the Rosetta mission, essentially for payload checkouts and functional tests. Housekeeping and science data were collected to check the instrument health, confirm that commands were successfully executed, and test new or revised flight procedures. 

The typical SESAME-PP health check generates telemetry on internal electronics reference voltages, the potentials at the two receiving sensors (i.e., $V_8$ and $V_{12}$), their difference ($\Delta V$), and the currents flowing through the three transmitter electrodes ($I_4$ at the +X sole, $I_{17}$ at the MUPUS-PEN hammering device and $I_{15}$ on the lid of the APXS sensor housing) to verify that these signals are not disturbed by noise and remain within the expected limits. These measurements were performed at various sounding and sampling frequencies. Because the landing gear is folded, the potentials measured during the cruise phase cannot be used for science nor reference purposes.

Tests were also conducted to examine the level of the interferences generated by CONSERT soundings and by Philaes’s flywheel during the SESAME-PP operations. It was discovered that the CONSERT signal strongly affects both SESAME-PP passive and active observations. CONSERT transmits an RF signal of $0.2$ s every $2.5$\,s, and it was first assumed that SESAME-PP could be operated between the radar pulses, which was later proven to be wrong (Sect. \ref{Section32}). It was recommended to  operate the flywheel with the lowest possible rotation rate after  the separation of Philae from Rosetta to limit electrical noise.

The only notable change in the SESAME-PP performance during the cruise was observed after the flyby of asteroid (2867) Steins (September 5, 2008): the level of the transmitted current in +X changed slightly at all frequencies. This is probably because of a change of stray parasite capacitances when the Rosetta spacecraft was rotated to examine the asteroid, thus exposing the Philae module to the Sun for about half an hour and likely changing the position of the electrode slightly with respect to the grounded structures. The transmitted current then remained stable until the end of cruise. The standard deviation of the currents measured before and after the Steins flyby gives a useful indication on the precision of the SESAME-PP measurements (namely, $28$\,nA on the amplitude and $0.9\ ^{\circ}$ on the phase).

The post-hibernation tests in March 2014 showed that all SESAME hardware and software had successfully survived the 31 months of hibernation of the Rosetta probe. It was also discovered that the RF link disturbed SESAME-PP, but it was anticipated that this interference would fade away after separation. 
 
Lastly, only passive measurements were performed during the predelivery phase in October 2014 to monitor the plasma environment over one full rotation of the comet and look for possible variations of the comet activities. These observations were conducted in cooperation with those of the Rosetta Plasma Consortium instruments RPC-MIP and RPC-LAP (Langmuir Probe). Both SESAME-PP and RPC (Lebreton, personal communication, 2016) detected no significant signature of dust impact or plasma effects besides some plasma wave activities during one of the measurements, indicating that a separation from the comet between $10$ and $15$\,km at the beginning of the descent was indeed suitable for a final calibration (Sect. \ref{Section32}).

\subsection{Separation, Descent, Landing  (SDL)}\label{Section32}
                
Four data blocks were acquired throughout the SDL phase that started on November 12, 2014 at $08:35:00$ UTC (Table \ref{table:4}): 
\begin{enumerate}[i)]
\item before separation,
\item immediately after separation with the landing gear deployed,
\item outside the Rosetta spacecraft zone of influence, and
\item shortly before nominal touchdown. 
\end{enumerate}

The SESAME-PP measurements acquired during the first sequence are in line with those performed during cruise. Only passive measurements were conducted during the second block for a joint RPC/SESAME-PP plasma environment monitoring. The third and fourth blocks were primarily dedicated to the calibration of the instrument in a near-vacuum environment. $I_4$, $V_8$, $V_{12}$ and $\Delta V$ were measured and acquired in the form of time series at $409$\,Hz and $758$\,Hz and processed onboard data (i.e., the phase and amplitude of $I_4$ and $\Delta V$) were acquired at these same frequencies and five additional frequencies, namely, $74$, $146$, $2946$, $6510,$ and $10080$\,Hz. These measurements were all the more crucial as no calibration could be performed with the flight model of Philae with deployed landing gear before launch.

\begin{figure}
\centering
\includegraphics[width=8.8cm]{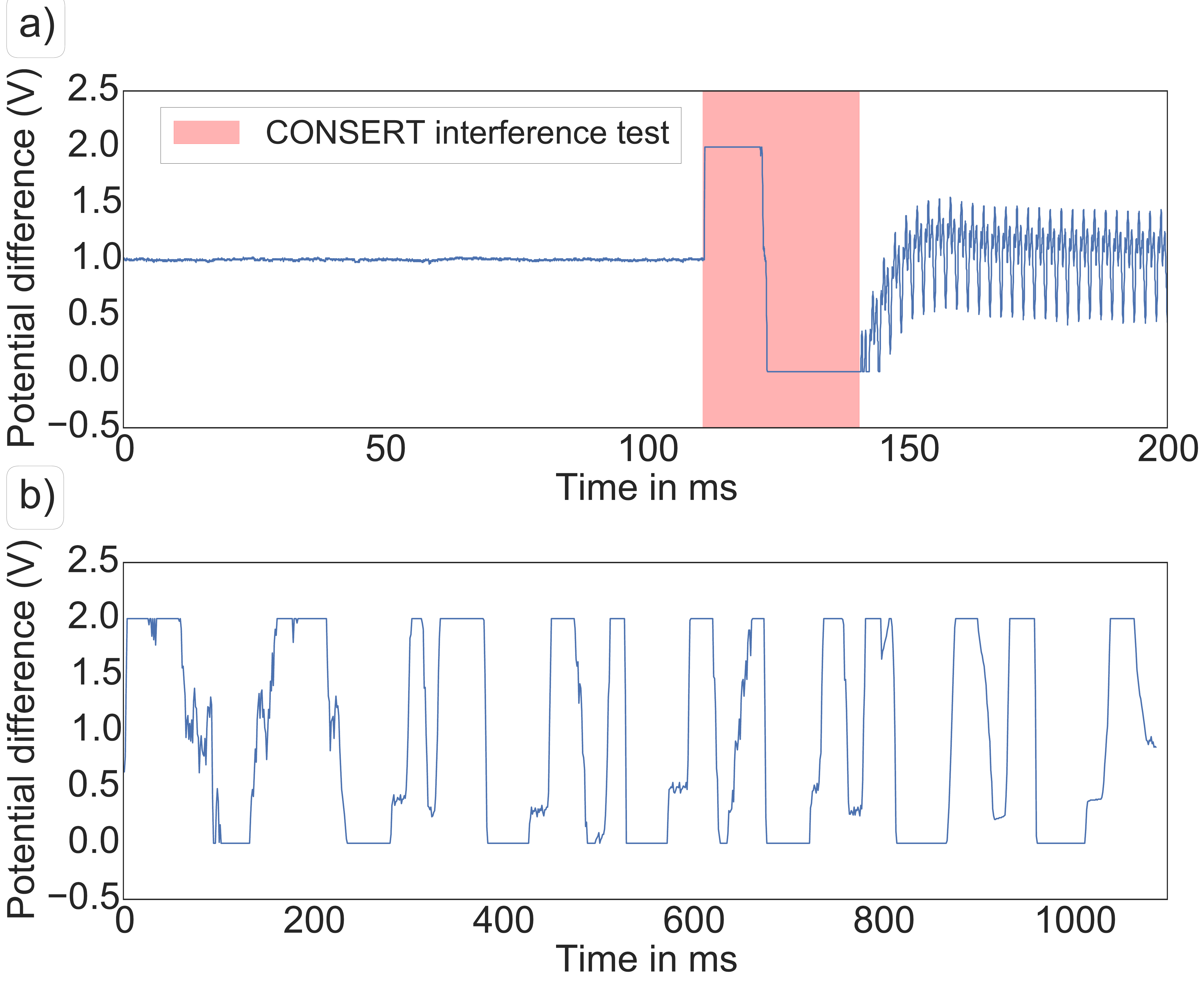}
\caption{Top panel: time series of the potential difference between the receiving electrodes during an interference test between SESAME-PP and CONSERT conducted in 2010. CONSERT operations clearly saturate the SESAME-PP receiving channels at $132$ ms. Bottom panel: time series of the potential difference between the receiving electrodes during the descent of the Philae lander toward the comet. The saturation pattern is similar to that observed during the SESAME-PP/CONSERT interference test.}
\label{Fig9}%
\end{figure}

Unfortunately, all potential measurements performed during the third and fourth block of the SDL phase were saturated (see Fig. \ref{Fig9} for an example). The observed disturbances are undoubtedly due to interferences generated by CONSERT sounding operations that stopped only three minutes after SESAME-PP’s last measurement block during the SDL phase. As a consequence, we do not have an in-flight calibration data from the instrument and, in particular, we are working under the assumption that the receiving electrodes and, specifically, the embedded preamplifiers have evolved similarly during Rosetta’s ten-year journey to the comet. This assumption is supported by the identical amplifier design and preselection of individual models with highly similar characteristics, as well as uniform exposure of both flight models to temperature, radiation, and vacuum. The transmitted currents on the +X foot, fortunately, were not perturbed and can be used for comparison to the currents measured at the surface of the comet during the FSS phase (see Sect. \ref{Section42}).

\subsection{First Science Sequence on the surface}\label{Section33}

On November 12, 2014, at 15:34:06 UTC, the harpoon ejection failed during the first touchdown of Philae on comet 67P/C-G. The lander bounced away from the surface before eventually coming down to a rest at its final landing site, Abydos, about two hours later (see \citet{Biele2015} for a description of the Philae multiple landings and bounces). The Philae module was then commanded to enter a “safe mode” that consists of four measurement blocks that require no mechanical activity e.g., DIM, COSAC and PTOLEMY sniffing, ROMAP, and MUPUS-TM (Thermal Mapper). These included SESAME-PP in a reduced geometry mode that only uses the three foot electrodes (+X as transmitter and --Y and +Y as receivers, see Sect. \ref{Section21}). The MUPUS-PEN and APXS were deployed on November 14, 2014, but could not be used as transmitters for SESAME-PP because of the constraints on Philae’s operations time. The lander entered hibernation on November 15, 2014 at 00:36:05 UTC when its batteries ran out due to the poor sunlight illumination of the Abydos site.
        
As a consequence, SESAME-PP did not operate in a nominal quadrupolar configuration during FSS. Instead, it performed four identical measurement blocks using the three foot electrodes; the lander body played the role of the fourth electrode. The blocks were performed on November 13, 2014 at two-hour intervals, starting shortly after local sunset and continuing into local night (Table \ref{table:4}). Each block consisted of one health check, two passive measurements and 11 active measurements. In the active mode, $I_4$, $V_8$, $V_{12}$, and $\Delta V$ were measured and collected in the form of time series at $409$\,Hz and $758$\,Hz, while only the phase and amplitude of $I_4$ and $\Delta V$ were acquired at $74$, $146$, $2946$, $6510,$ and $10080$ Hz. The active measurements of each block lasted for about 2 min.

The first measurement block (FSS1) started shortly after the end of the sunlit period with +X and +Y feet at a temperature of about $-120\ ^{\circ}$ C, while the --Y foot temperature, permanently in shadow, was at $-165\ ^{\circ}$C. By the time of the fourth measurement block (FSS4), all three feet had reached the same temperature of about $-165\ ^{\circ}$ C, which enables the temperature variation effects to be monitored. The quoted temperatures were measured by the PT1000 thermal sensors attached to the SESAME-CASSE sensors and may differ from the real temperatures of the +Y and --Y preamplifiers. In particular, the PT1000 sensors are insulated from the close environment with a guarding kapton-aluminium foil, while the SESAME-PP receivers are in good thermal contact with the lid of the soles. As a consequence, it is likely that SESAME-PP receivers cooled faster after sunset than indicated by the SESAME-CASSE temperature sensors. Moreover, while the accuracy of the SESAME-CASSE temperature sensors is $\pm 2\ ^{\circ}$C at $-100\ ^{\circ}$C, it is $\pm 10\ ^{\circ}$C around and below $-160\ ^{\circ}$C. Lastly, we note that, prior to the launch of the Rosetta probe, SESAME-PP preamplifiers had not been calibrated for temperatures below $-150\ ^{\circ}$C. Additional tests have been conducted at LATMOS after the Philae landing on ground models of the preamplifiers in a cryogenic room for temperatures down to $-175\ ^{\circ}$C to provide a more reliable calibration of the data against temperature.  

\begin{figure}
\centering
\includegraphics[width=8.8cm]{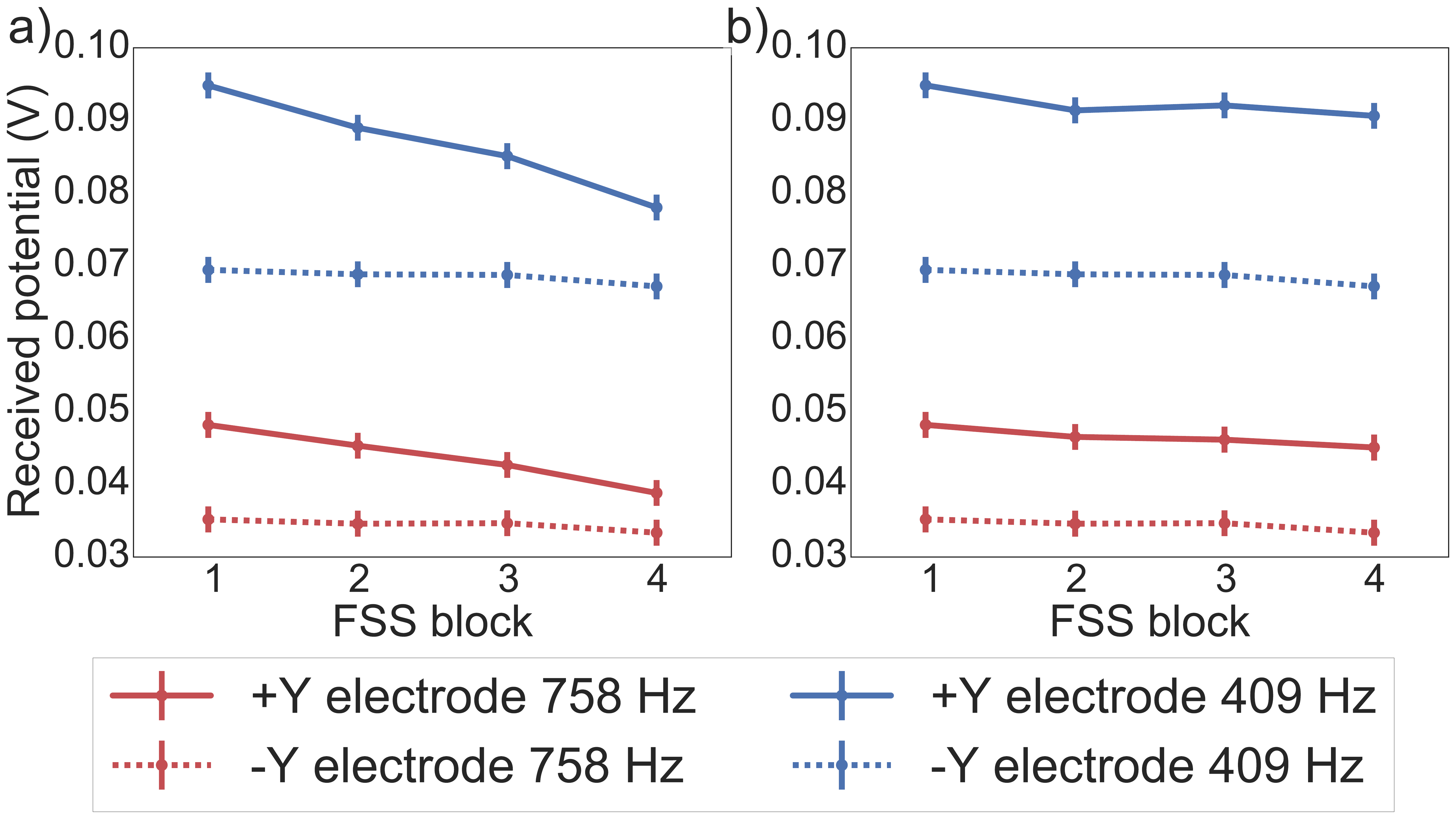}
\caption{Calibrated measured potentials on the SESAME-PP receiving electrodes --Y and +Y for the four measurement blocks of the FSS. a) The temperatures used for calibration are those measured by the SESAME-CASSE PT1000 sensors.  The potential of the +Y electrode decreases with time. b) The temperatures used for the calibration for the --Y electrode and the FSS4 measurements of the +Y electrode are those measured by the SESAME-CASSE PT1000 temperature sensors, minus $10\ ^{\circ}$C. The drop in potential of the +Y electrode has now disappeared.}
\label{Fig10}
\end{figure}

As shown in Fig. \ref{Fig10}a, a drop in potential was observed throughout the night at the +Y foot, while the potential measured on --Y remained constant. The data were corrected for the temperature dependence of the preamplifier gain using the latest calibration and temperatures indicated by SESAME-CASSE sensors. However, we cannot rule out that this drop could be due to an incomplete correction of the temperature dependence of the electronics. As a matter of fact, if the +Y receiver was $10\ ^{\circ}$C cooler than indicated by the SESAME-CASSE temperature sensor during the fourth block and if the --Y foot was $10\ ^{\circ}$C cooler during the entire FSS, which is possible given the accuracy of the thermal sensors below $-150\ ^{\circ}$C and the measurements of the MUPUS-TM instrument with reference to temperatures in the range $-183\ ^{\circ}$C to  $-143\ ^{\circ}$C \citep{Spohn2015}, then the trend would have disappeared once the appropriate calibration correction was applied (Fig. \ref{Fig10}b). As a further argument, the electrical properties of water ice and other potential candidates for the surface material of the nucleus are not expected to vary at such low temperatures (see Sect. \ref{Section13}). 

The most interesting feature shown in Fig. \ref{Fig10} is the significant potential difference measured between the two receiving electrodes: the potential of +Y is significantly larger than that of --Y with a ratio $V_{12}/V_{8}$  of $1.35\pm 0.03$. In a homogeneous environment the potential of the two feet are expected to be nearly identical (i.e., a ratio $V_{12}/V_{8}$  of $1.08$ would be expected owing to the slight asymmetry induced by the body rotation of $11.2\ ^{\circ}$ relative to the landing gear). The potential difference was observed both at $409$ and $758$ Hz. We estimate that this feature is genuine and not an effect of any hypothetical temperature gradient since it was still present at the end of the night (i.e., during FSS4) when both feet were at the same temperature. Strictly speaking, in the absence of any calibration during the SDL, we cannot completely rule out a drift of one of the receiving channels with respect to the other during the long journey to the comet. There is however no direct evidence for such a drift, and given the identical preamplifier and electronics design, near-identical prelaunch characteristics, and identical environment during cruise, an identical drift, if any, can be assumed. In addition we recall that no drift was observed on PWA-HASI/Huygens after the seven-year cruise of the Cassini–Huygens probe to the Saturnian system. The difference between the potentials of the --Y and +Y electrodes is further analyzed and compared to numerical simulations in Sect. \ref{Section43}. 

Lastly, the amplitude of transmitted current on foot +X is analyzed in light of the current measured during the descent phase in Sect. \ref{Section42}.

\begin{table*}
\centering
\caption{SESAME-PP active measurements during the SDL and FSS phases.}
\label{table:4}
\begin{tabular}{l l l l l}
\hline
Measurement block\tablefootmark{a}&Date&Start-End (UTC)&\begin{tabular}[c]{@{}l@{}}Frequency (Hz)\tablefootmark{b}\end{tabular}                               & \begin{tabular}[c]{@{}l@{}}Foot temperatures ($^{\circ}$C)\\ +Y (top) --Y (bottom)\tablefootmark{c}\end{tabular}   \\ \hline
SDL1              & 2014-Nov-12  & 07:41:01-07:41:57 & TS: 758, 409                                                                           & \begin{tabular}[c]{@{}l@{}} -135.0\\ - 140.8\end{tabular}  \\ \hline
SDL3              & 2014-Nov-12 & 09:05:04-09:11:21 & TS: 758, 409                                                                           & \begin{tabular}[c]{@{}l@{}} -133.9\\ - 138.8\end{tabular}  \\ \hline
SDL4              & 2014-Nov-12 & 14:47:23-14:48:30 & TS: 758, 409                                                                           & \begin{tabular}[c]{@{}l@{}} -124.3\\ - 125.2\end{tabular}  \\ \hline
FSS1              & 2014-Nov-13 & 08:10:49-08:13:07 & \begin{tabular}[c]{@{}l@{}}TS: 758, 409\\ PD: 10080, 6510, 2948, 146, 74\end{tabular}  & \begin{tabular}[c]{@{}l@{}} -131.8\\ - 161.8\end{tabular}  \\ \hline
FSS2              & 2014-Nov-13 & 10:12:47-10:15:07 & \begin{tabular}[c]{@{}l@{}}TS: 758, 409\\ PD: 10080, 6510, 2948, 146, 74\end{tabular}  & \begin{tabular}[c]{@{}l@{}} -145.7\\ - 162.1\end{tabular}  \\ \hline
FSS3              & 2014-Nov-13 & 12:14:09-12:17:09 & \begin{tabular}[c]{@{}l@{}}TS : 758, 409\\ PD: 10080, 6510, 2948, 146, 74\end{tabular} & \begin{tabular}[c]{@{}l@{}} -156.8\\ - 163.4\end{tabular}  \\ \hline
FSS4              & 2014-Nov-13 & 14:16:51-14:19:11 & \begin{tabular}[c]{@{}l@{}}TS: 758, 409\\ PD: 10080, 6510, 2948, 146, 74\end{tabular}  & \begin{tabular}[c]{@{}l@{}} -161.9\\  - 164.1\end{tabular} \\ \hline
\end{tabular}

\tablefoot{
\tablefoottext{a}{The SDL2 block was only dedicated to passive measurements and is therefore not shown here}
\tablefoottext{b}{TS stands for “time series” and PD for “onboard processed data”.}
\tablefoottext{c}{as measured by SESAME-CASE PT1000 temperature sensors}
}
\end{table*}

\section{Analysis of the SESAME-PP\ surface data}\label{Section4}

\subsection{Approach}\label{Section41}

As described earlier, the most interesting observation at Abydos is the significant potential difference measured between the two receiving electrodes. Although the absence of calibration during the descent phase leaves a theoretically possible drift of one channel with respect to the other, we assume that both receiving electronic circuits aged in the same way and we investigate the implications of this observation in terms of electrical properties and distribution of the matter around the SESAME-PP electrodes, assuming a homogenous composition. 

Our approach consists in comparing the flight model measurements to a number of numerical simulations based on realistic models, using the Capacity-Influence Matrix Method described in Sect. \ref{Section15}. This method requires a good knowledge of the configuration of operation at Abydos which was far from nominal. As a matter of fact, the landing of Philae did not take place as planned. Philae bounced several times until it reached Abydos where it came to a rest in what looks like a cavity just slightly larger than the size of the lander, and partially shadowed by nearby boulders or cliffs with the +Y foot pointing upward, the --Y foot pointing downward, and the +X foot close to or resting on a Sun-illuminated surface \citep{Bibring2015}.

Section \ref{Section431} is dedicated to the reconstruction of the attitude and environment of SESAME-PP at Abydos using almost all available constraints, including those provided by the measurements of the current transmitted on the +X foot (analyzed in Sect. \ref{Section42}). We emphasize that if Philae had come to a rest in a nominal horizontal position, resting on its three legs, in the absence of measurements that employed the other transmitting electrodes, we would have most likely observed no meaningful difference of potential between the two receiving electrodes because the configuration of operation would have been almost perfectly symmetrical (hence, $\Delta V \approx 0$, see Eq. \ref{Eq16}). In this regard, the acrobatic attitude of Philae at Abydos offers an opportunity for SESAME-PP to provide insights into the near surface of 67P/C-G at this location. 

\subsection{The transmitted currents}\label{Section42}

\begin{figure*}
\centering
\includegraphics[width=18cm]{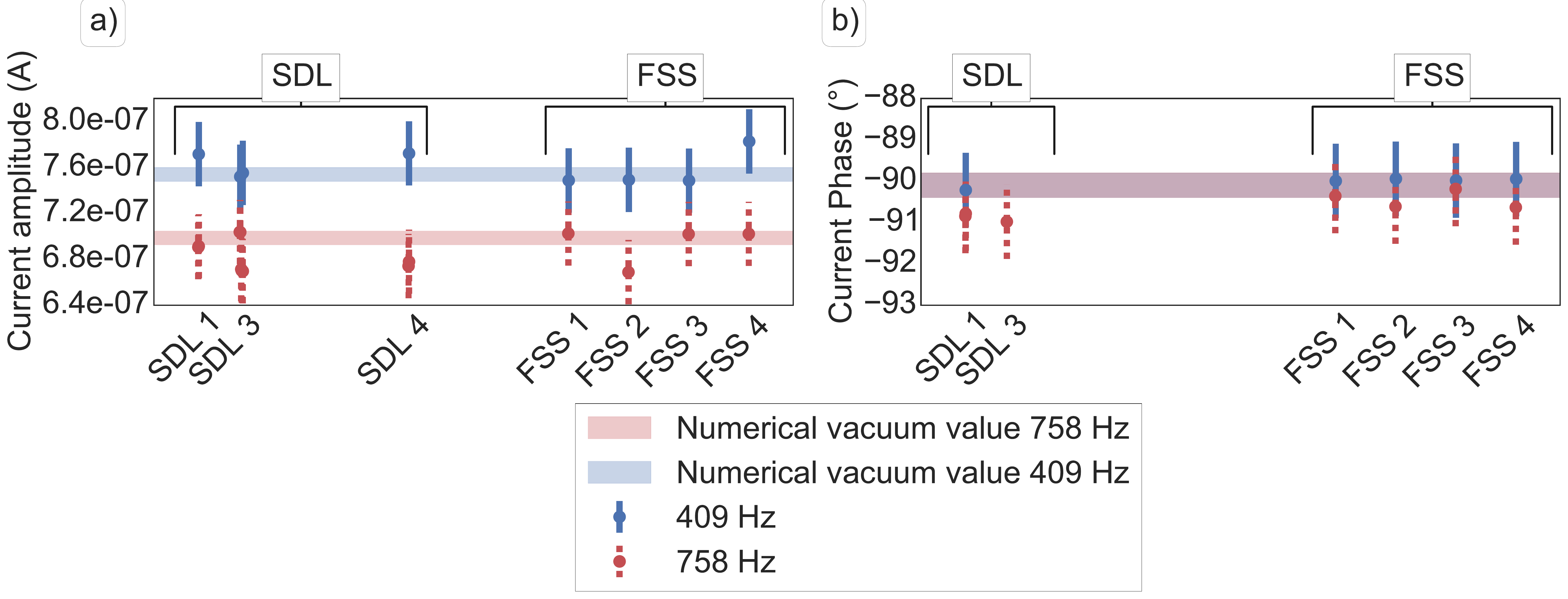}
\caption{Dots with error bars represent the amplitude (a) and phase (b) of the currents measured during SDL and FSS phases at $409$\,Hz and $758$\,Hz. The phase of the currents measured during SDL 4 could not be retrieved because of disturbances in the signal. The blue (resp., red) shaded lines indicates the expected amplitude of the current and phases at $409$\,Hz (resp., $758$\,Hz) in a vacuum derived from numerical simulations and the associated error due to the mesh approximation. The potential difference used to generate the current at $758$\,Hz is half that used at $409$\,Hz}
\label{Fig11}
\end{figure*}

A potential difference is applied between the sole of the transmitting electrode and the body, generating a current that depends upon the electrical properties of the environment around the transmitting electrode. The current injected through the +X electrode was measured (amplitude and phase) both during the SDL phase, in a near-vacuum environment, and at the beginning of the FSS phase, on the surface of 67P/C-G nucleus. No significant difference (within the error bars) is noted between these two sets of measurements. Fig. \ref{Fig11} also shows that both SDL and FSS are close to the value expected in vacuum from numerical simulations (Eq. \ref{Eq24}). This observation suggests that, during FSS, the transmitting electrode (+X) was not in contact with the nucleus surface and/or that the material under the transmitting electrode has very low dielectric constant and conductivity.

The SESAME-CASSE data recorded during the hammering of MUPUS-PEN contain information on the quality of the contact between the Philae feet (where CASSE sensors are also located) and the surface. They show that, at the beginning of the hammering session, only the +Y foot reliably recorded MUPUS-PEN strokes; the sensors in the +X and --Y feet detected them at a later stage \citep{Knapmeyer2016}. This suggests that Philae moved during the initial phase of the session, eventually settling in an attitude that improved the coupling between the foot sensors and the surface. Because SESAME-PP measurements were performed before the deployment of the MUPUS boom, this would imply that the contact between the transmitting +X electrode and the surface was bad or even nonexistent when the transmitted currents shown on Fig. \ref{Fig11} were measured.  More specifically, a distance of only 1\,cm between the +X foot and the surface would be sufficient to explain the absence of a significant difference between the currents measured during SDL and FSS.

However, if we assume that the +X foot was close to the surface ($<1$\,cm), SESAME-PP current measurements place a constraint on the upper limit of the dielectric constant and conductivity of the surrounding material. Comparison with numerical simulations yields a maximum dielectric constant of $3$ and a maximum conductivity of $4\cdot 10^{-8}$\,S/m. With such a low conductivity, the surface material of the nucleus could be regarded as a pure dielectric.  

\subsection{The received potentials }\label{Section43}

As mentioned earlier, to retrieve the permittivity of the surface from the comparison of the measured received potentials and the numerical simulations derived from the Capacity-Influence Matrix Method, a good knowledge of the configuration and environment of SESAME-PP operations is required. Almost all available constraints were thus gathered to build a suite of realistic and reliable geometry models of the environment and attitude of Philae at Abydos. These models were first constructed, using the free and open-source Blender software, and then imported into COMSOL Multiphysics™  to simulate the SESAME-PP operations numerically. Unfortunately, the uncertainties on the phases of the received potentials are too large to estimate the electrical conductivity of the near surface; the dielectric constant can however be retrieved.

\subsubsection{Reconstruction of Philae attitude and environment at Abydos during SESAME-PP operations }\label{Section431}

\begin{figure}
\centering
\includegraphics[width=8.8cm]{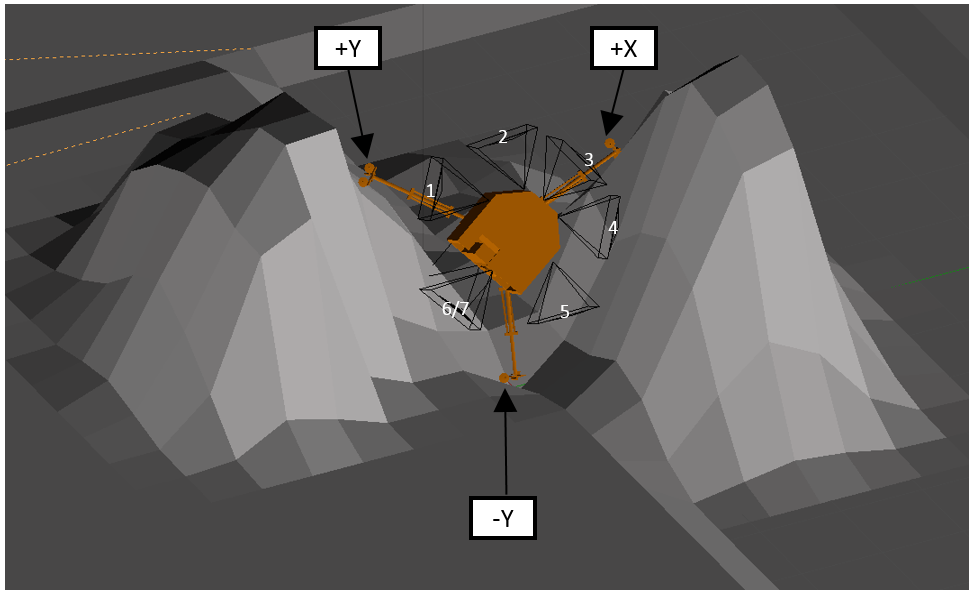}
\caption{Possible 3D model of the Philae attitude and environment at Abydos. The fields of view and numbers of the CIVA cameras are indicated with black triangles. The model was built using the Blender software (http://www.blender.org/)}
\label{Fig12}
\end{figure}

\begin{figure}
\centering
\includegraphics[width=8.8cm]{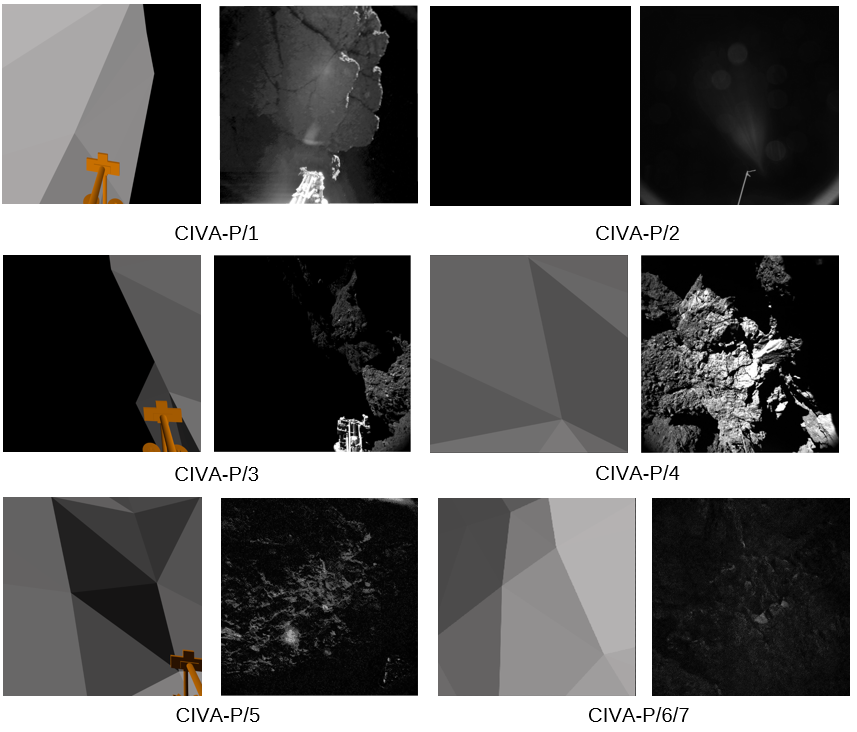}
\caption{Comparison between the CIVA images (credits: ESA/Rosetta/Philae/CIVA) taken at Abydos and the corresponding camera views in the model of Philae environment and attitude built under Blender. }
\label{Fig13}
\end{figure}

In order to reconstruct the attitude and environment of Philae at its final landing site, we took into account constraints from various origins:

\begin{enumerate}
\item \emph{Comet Infrared and Visible Analyzer (CIVA) images} \citep{Bibring2015}:
The CIVA panorama at Abydos consists of a set of seven images around the Philae body taken by cameras with well-known positions and fields of view (Fig. \ref{Fig12} and Fig. \ref{Fig13}) and provides a wealth of constraints for the reconstruction of the attitude and surroundings of the lander. Revealing that one of the $693$\- mm long CONSERT antenna is touching the surface, camera 3 even gives a quantitative indication of the distance of the “walls” of the hole in which Philae rested. In addition, a pair of stereo images taken by cameras 5 and 6 in the direction of the lander balcony allows us to evaluate distances (in the range $80$\,cm-$7$\,m) and reconstruct the 3D environment in this direction. 
\item \emph{ROsetta Lander Imaging System (ROLIS) images} (S. Mottola, personal communication):
Pointing under the lander, the ROLIS instrument provides two additional images as well as distance information by stereography.
\item \emph{MUPUS} \citep{Spohn2015}:
Additional constraints can be obtained from two of the three MUPUS instruments. First, MUPUS-TM detected direct illumination behind the lander in the direction of the PEN deployment, which completes information from the solar array telemetry (these are not taken into account in the present model). Second, the MUPUS-PEN probe was nominally deployed and started the hammering sequence. While it is not clear whether or not the probe hit an obstacle during its $58.5$\,cm long deployment \citep{Spohn2015}, the $30$\,mm long MUPUS-PEN probe most likely touched the surface without fully penetrating it.
\item \emph{SESAME-CASSE} \citep{Knapmeyer2016}:
As mentioned previously SESAME-CASSE recorded the hammering of the MUPUS-PEN and the clear signal that was measured by the +Y accelerometers strongly suggests that this foot was in good contact with the surface. On the --Y and +X feet, the signal was weak at the beginning of the hammering sequence and then increased as if the lander had slightly moved, thereby enhancing the contact between this foot and the “ground”. We note that SESAME-PP operation occurred before the MUPUS-PEN hammering and therefore at a time when the contact of the +X and --Y electrodes with the surface was possibly poor, even nonexistent.
\item \emph{SESAME-PP}:
Though not very constraining, SESAME-PP current measurements suggest that the +X foot is not necessary resting on the surface (see Sect. \ref{Section42}); this is consistent with SESAME-CASSE  first measurements on the +X accelerometer. Further, the ratio between the potentials measured on the +Y and --Y feet (i.e., $V_{12}/V_8=1.35\pm 0.03$) suggests that the +Y foot may be surrounded by and/or closer to a greater amount of cometary material than the --Y foot. 
\end{enumerate}

The geometry model of Philae attitude and environment at Abydos has two main degrees of freedom. First, the amount of matter located under the +Y foot in the blind spot of the CIVA and ROLIS cameras. If a small amount of matter is present there, then a high subsurface dielectric constant is required to reconcile simulations with SESAME-PP data and, in particular, to retrieve the measured ratio $V_{12}/V_8$.  Second, the quality of the contact between SESAME-PP receiving electrodes and the surface controls the measured potentials. The better this contact is, the higher the received potential.

Figure \ref{Fig12} shows a possible 3D model of the Philae lander attitude and environment at Abydos. This model satisfies all of the constraints previously listed and was built to provide a lower bound for the subsurface dielectric constant. This was carried out by adding as much cometary material as possible under the +Y foot, and giving +Y (respectively, --Y) a good (respectively, bad) contact with the surface. The simulated CIVA panorama obtained with this 3D model is compared to the actual panorama taken by CIVA in Fig. \ref{Fig13}.

\subsubsection{Retrieval of the dielectric constant of the near surface of Abydos}\label{Section432}

The model of the Philae lander attitude and environment at Abydos presented in Fig. \ref{Fig12} was found after varying the two degrees of freedom mentioned in Sect. \ref{Section431} to retrieve the lowest possible dielectric constant of the subsurface. Applying the Capacity-Influence Matrix Method to this model, we compute the ratio between the potential amplitude of the two receiving feet (i.e., $V_{12}/V_8$) varying the dielectric constant of the subsurface around Philae from 1 to 5 and setting the conductivity to zero. We find that the value for which simulation best reproduces SESAME-PP observations is $2.45\pm 0.20$.
 
We emphasize that this value is a strict lower limit: any other geometrical model satisfying the constraints listed in Sect. \ref{Section431}, but with less material around +Y and/or a better contact between the -Y electrode and the surface, requires a higher dielectric constant to be reconciled with SESAME-PP observations at Abydos. Furthermore, we verify that including a non-null conductivity results in a higher lower limit for the dielectric constant.

\section{Discussion and conclusions}\label{Section5}

\begin{figure*}
\includegraphics[width=18cm]{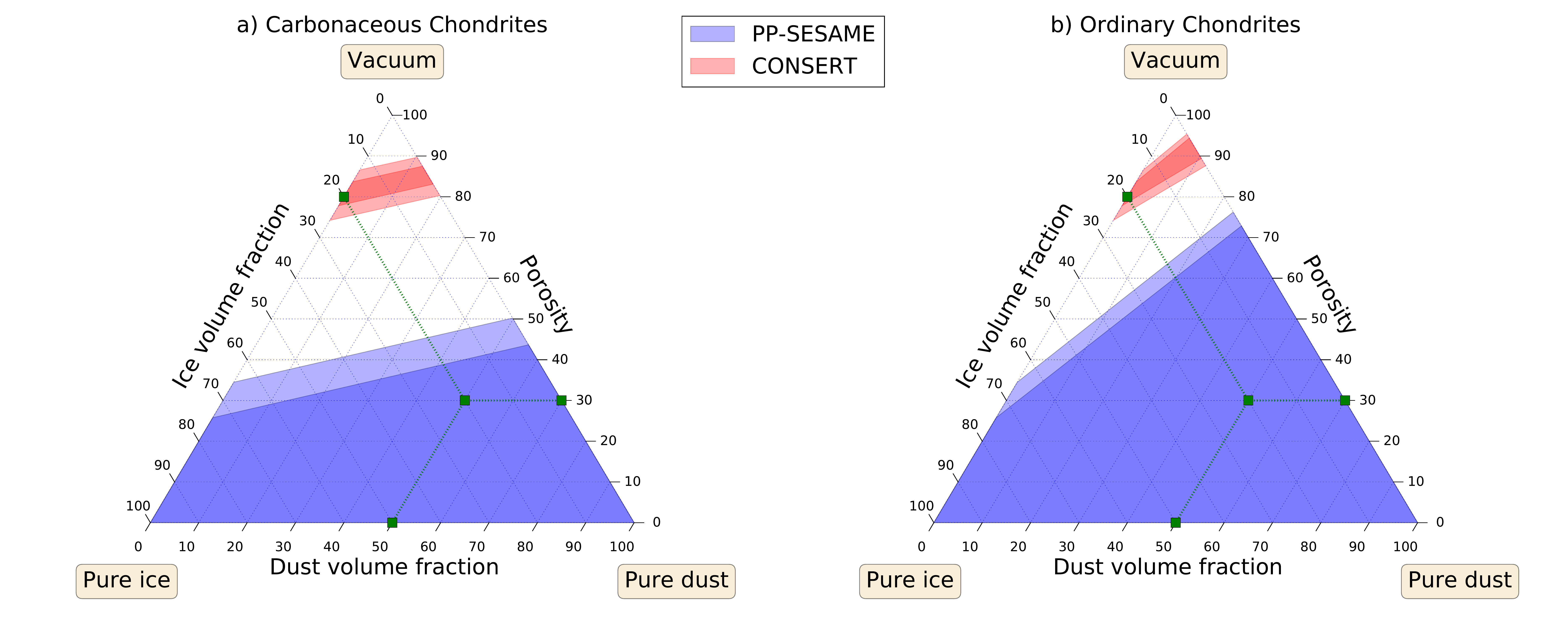}
\caption{Ternary diagram of dust, ice, and porosity volumetric fraction as derived from SESAME-PP (blue domain) and CONSERT (red domain) observations for carbonaceous (a) and ordinary (b) chondrites. The results from SESAME-PP are representative of the composition and porosity of the first meter of the 67P/C-G nucleus, while the results from CONSERT is an average for the first hundreds of meters of the interior of the small lobe of the comet. To help read this diagram, an example is shown (green squares) on the figure, corresponding to $30\%$ porosity, $50\%$ dust, and $20\%$ ice. The region with a lighter blue or red color show the error on the inferred dielectric constants.}
\label{Fig14}
\end{figure*}
                
The SESAME-PP measurements at the surface of 67P/C-G provide a lower limit for the dielectric constant of the near surface of the nucleus at Abydos in the frequency band of $409$\,Hz to $758$\,Hz. The inferred value of $2.45 \pm 0.20$ applies to the first meter of the nucleus (see Sect. \ref{Section23}). This result can be compared to investigations of the electrical properties of the nucleus with radars \citep{Kamoun2014, Kofman2015} and to laboratory measurements on putative cometary analogs \citep{Heggy2012}. It thus can bring new constraints on the porosity and composition of subsurface of the 67P/C-G nucleus and on their variations with depth.

Based on observations with the radar system of the Arecibo Observatory ($2.38$\,GHz) during a close encounter of the comet with the Earth, \citet{Kamoun2014} constrain the dielectric constant of the top $\sim 2.5$\,m of the subsurface of 67P/C-G to be in the range $1.9-2.1$ at that frequency. We emphasize that this value is an average for the whole surface of the nucleus. More recently, using the propagation time of the CONSERT signals in the upper part of the smaller lobe of 67P/C-G, \citet{Kofman2015} find that the average dielectric constant of the interior is very small, namely $1.27$, at the CONSERT operation frequency of $90$\,MHz.

Assuming that the surface material consists of a ternary mixture composed of a dust phase, an ice phase, and vacuum, and using a mixing law, the Arecibo, CONSERT, and SESAME-PP inferred dielectric constant can be used to invert the volumetric fraction of each of these phases at their respective sounding depths. Using the Hashin-Shtrikman bounds \citep{Sihvola1999} derived from Maxwell Garnett mixing formula (see Eq. \ref{Eq9}), \citet{Kofman2015} thus estimate that their result is consistent with a volumetric dust/ice ratio of $0.4$ to $2.6$ and a porosity of $75$ to $85\%$ (even higher for ordinary chondrites), while \citet{Kamoun2014} constrain the porosity of the first meters to be $\sim 70\%$. We follow strictly the same approach as in \citet{Kofman2015} for the inversion of the SESAME-PP derived dielectric constant.

For their analysis, \citet{Kofman2015} consider that the dust phase must be chondritic in nature. In absence of information on the dielectric constant of carbonaceous and ordinary chrondrites at the low frequencies of SESAME-PP, we use the same values as in \citep{Kofman2015}, namely values in the range $2.6-2.9$ for carbonaceous chondrites and 4.8--5.6 for ordinary chrondrites as measured by \citep{Heggy2012} on meteoritic samples. To support this assumption we note that geological materials at very low temperature (the temperature
at Abydos during SESAME-PP measurements ranges between $-165\ ^{\circ}$C
and $-130\ ^{\circ}$C; see Table \ref{table:4}) have relatively little variation
in the real part of the permittivity with frequency. In practice, we only use the upper bounds of these ranges ($2.9$ for carbonaceous chondrites and $5.6$ for ordinary chrondrites) since SESAME-PP observations provide a constraint only on the lower bound of the dielectric constant. We further note that the dielectric constant of chrondrites was measured for pellets with a porosity of $30\%$ so that the dust volumetric fraction contains $30\%$ of vacuum, which has to be taken into account. Regarding the ice phase, we use the highest value assumed by \citep{Kofman2015}, namely $3.1$, which corresponds to $100\%$ water ice. We emphasize that there is no approximation in using this value at low frequencies since, as previously mentioned in Sect \ref{Section13}, the dielectric constant of water ice at cryogenic temperatures loses its frequency dependence.

The constraints, in terms of dust-to-ice ratio and porosity, derived from SESAME-PP are presented on the ternary diagram in Fig. \ref{Fig14} (blue domain) next to CONSERT results (red domain). We note that SESAME-PP-derived constraints apply to the first meter of the near surface, while CONSERT-derived constraints apply to hundreds of meters below the surface.

The SESAME-PP results suggest that the first meter of the nucleus is more compacted with a porosity below $50\%$ for carbonaceous chondrites and below $75\%$ in the case of less primitive ordinary chondrites, than its interior as sensed by CONSERT. Though less constrained, a comparison between SESAME-PP and Arecibo results further suggests that there may also be a gradient in porosity in the first meters of the cometary mantle. 

The presence of a relatively resistant “shell” is supported by observations from the MUPUS instrument package that reveal that both the thermal inertia and surface strength at Abydos are larger than expected \citep{Spohn2015}. As mentioned before (Sect. \ref{Section431}), the MUPUS-PEN thermal probe could barely penetrate the near surface, pointing to a local resistance of at least 2 MPa, and the thermal inertia was found to be $85 \pm 35$ Jm\textsuperscript{-2}K\textsuperscript{-1}s\textsuperscript{-1/2}, which is consistent with a near-surface porosity in the range $30-65\%$ and most likely in the range $40-55\%$. At a larger scale ($>10$\,m), the idea of enhanced compaction near the surface (consequent to an increasing porosity with depth) is supported by the CONSERT data acquired at grazing angles that are consistent with a decreasing dielectric constant with depth \citep{Ciarletti2015}, although this latter result could also be attributed to a decreasing dust-to-ice ratio.

SESAME-PP observations place no constraint on the dust-to-ice ratio in the first meter below the surface. However, the dust abundance relative to ice is most likely much larger near the surface than deeper. Most geological models \citep[see][]{Belton2007} even predict a desiccated outer dust layer, possibly few meters thick, as a by-product of ice sublimation. If proven to be true, this would imply that the first meter of the subsurface consists of $55\%$ of dust and $45\%$ of porosity in the case of carbonaceous chondrites and $25\%$ of dust and $75\%$ of porosity in the case of ordinary chondrites (see Fig. \ref{Fig14}).  
  
On the other hand, MUPUS results suggest that the desiccated dust layer is thin at Abydos and, together with the SESAME-PP finding of an enhanced compaction of the near surface of the comet, suggest that some cementing processes are at play. These processes most likely involve ice that may sinter at each perihelion and refreeze as soon as the comet is receding from the Sun. The landing site of Philae was poorly illuminated in November 2014 and frozen ice was probably still present in the upper layers of the surface. 
 
As a further argument, the appearance of Abydos is consolidated as defined by \citet{El-Maarry2015} and the network of fractures revealed by CIVA images on the walls of the cavity in which Philae settled (see Fig. \ref{Fig13}), which are also seen at a larger scale by the OSIRIS Rosetta camera, is often an indication of the contraction of ice below the surface \citep{El-Maarry2015a}. In addition, CIVA images show variations in the surface reflectance at cm down to mm scale; the brighter spots in the observed granular grains could be ice-rich \citep{Bibring2015}.

\begin{acknowledgements}
Rosetta is a Cornerstone Mission of ESA and and the Philae lander is provided by a consortium led by Deutsches Zentrum f\"ur Luft und Raumfahrt (DLR), Max Planck Institut f\"ur Sonnensystemforschung (MPS), Centre National d'Etudes Spatiales (CNES), and Agenzia Spaziale Italiana (ASI). SESAME is an element of the Rosetta lander Philae payload. It consists of three instruments, CASSE, DIM, and PP, which were provided by a consortium comprising DLR, MPS, FMI, MTA EK, Fraunhofer IZFP, Univ. Cologne, LATMOS, and ESA/ESTEC. The authors are grateful to CNES, and in particular to Philippe Gaudon, for supporting the SESAME-PP experiment. They also thank Eric Jurado (CNES) for leading a working group to create a digital terrain model of the environment of Philae at Abydos and the members of this group: Valentina Lommatsch, Felix Finke, and Jörg Knollenberg from DLR and Cedric Delmas, Romain Garmier, Emile Remetean, and Alex Torres from CNES. They are also grateful to Stefano Mottola for sharing the ROLIS images taken at Abydos before their publication and to Alain Hérique from the CONSERT team for clarifiyng the formulae used in the ternary diagrams shown in Kofman et al. (2015). At ESA/ESTEC, we thank Bengt Johlander and Bart Butler for excellent technical and management support provided during the instrument development phase.  AL is supported by the Région Ile-de-France (DIM-ACAV). ALG benefits from a Chair at the French Space Agency CNES/UVSQ (Université Versailles Saint-Quentin). 

\end{acknowledgements}

\bibliographystyle{aa} 
\bibliography{AA201628304} 

\begin{appendix}
\section{Additional figures}

\begin{figure*}
\includegraphics[width=16.4cm,clip]{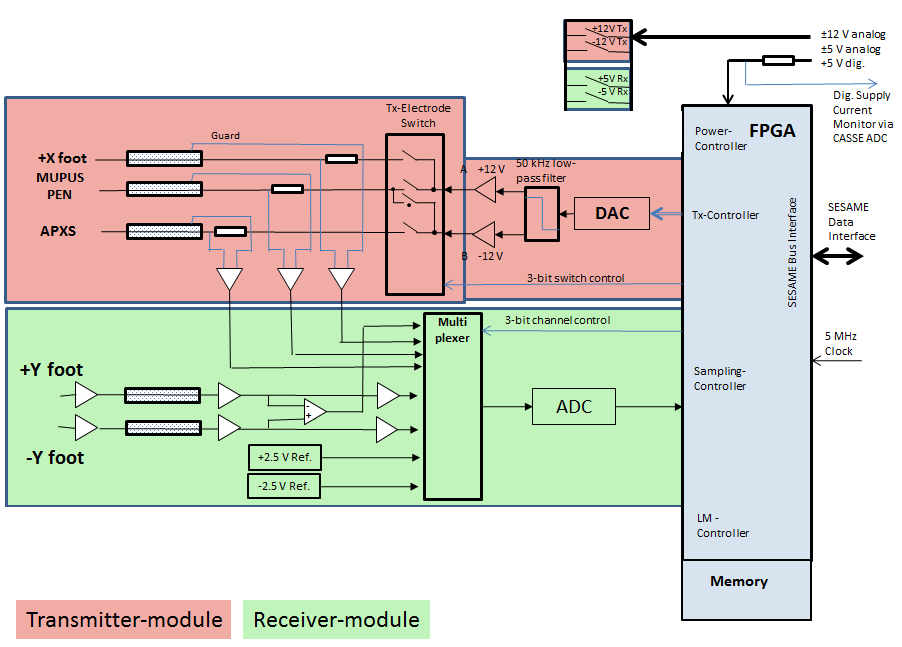}
\caption{SESAME-PP instrument block-diagram}
\label{Fig4}
\end{figure*}

\section{Additional tables}

\longtab{
\onecolumn
\begin{landscape}
\begin{longtable}{llllllllllllllllllllllllllllllllllllllllll}
\caption{Real (top) and imaginary (bottom) part of the electronic capacitance matrix $\vec{[K^\mathrm{e}]}$(pF).}\\
\hline\hline
El. & 1   & 2 & 3      & 4      & 5   & 6 & 7     & 8 & 9   & 10 & 11    & 12 & 13     & 14 & 15 & 16     & 17     & 18     & 19     \\\hline
1       & -21 & 0 & 21     & 0      & 0   & 0 & 0     & 0 & 0   & 0  & 0     & 0  & 0      & 0  & 0  & 0      & 0      & 0      & 0      \\
2       & 0   & 0 & 0      & 0      & 0   & 0 & 0     & 0 & 0   & 0  & 0     & 0  & 0      & 0  & 0  & 0      & 0      & 0      & 0      \\
3       & 21  & 0 & -95400 & 93300  & 0   & 0 & 0     & 0 & 0   & 0  & 0     & 0  & 2100   & 0  & 0  & 0      & 0      & 0      & 0      \\
4       & 0   & 0 & 93300  & -93300 & 0   & 0 & 0     & 0 & 0   & 0  & 0     & 0  & 0      & 0  & 0  & 0      & 0      & 0      & 0      \\
5       & 0   & 0 & 0      & 0      & -21 & 0 & 21    & 0 & 0   & 0  & 0     & 0  & 0      & 0  & 0  & 0      & 0      & 0      & 0      \\
6       & 0   & 0 & 0      & 0      & 0   & 0 & 0     & 0 & 0   & 0  & 0     & 0  & 0      & 0  & 0  & 0      & 0      & 0      & 0      \\
7       & 0   & 0 & 0      & 0      & 21  & 0 & -2120 & 0 & 0   & 0  & 0     & 0  & 2100   & 0  & 0  & 0      & 0      & 0      & 0      \\
8       & 0   & 0 & 0      & 0      & 0   & 0 & 0     & 0 & 0   & 0  & 0     & 0  & 0      & 0  & 0  & 0      & 0      & 0      & 0      \\
9       & 0   & 0 & 0      & 0      & 0   & 0 & 0     & 0 & -21 & 0  & 21    & 0  & 0      & 0  & 0  & 0      & 0      & 0      & 0      \\
10      & 0   & 0 & 0      & 0      & 0   & 0 & 0     & 0 & 0   & 0  & 0     & 0  & 0      & 0  & 0  & 0      & 0      & 0      & 0      \\
11      & 0   & 0 & 0      & 0      & 0   & 0 & 0     & 0 & 21  & 0  & -2120 & 0  & 2100   & 0  & 0  & 0      & 0      & 0      & 0      \\
12      & 0   & 0 & 0      & 0      & 0   & 0 & 0     & 0 & 0   & 0  & 0     & 0  & 0      & 0  & 0  & 0      & 0      & 0      & 0      \\
13      & 0   & 0 & 2100   & 0      & 0   & 0 & 2100  & 0 & 0   & 0  & 2100  & 0  & -10500 & 0  & 0  & 2100   & 0      & 2100   & 0      \\
14      & 0   & 0 & 0      & 0      & 0   & 0 & 0     & 0 & 0   & 0  & 0     & 0  & 0      & 0  & 0  & 0      & 0      & 0      & 0      \\
15      & 0   & 0 & 0      & 0      & 0   & 0 & 0     & 0 & 0   & 0  & 0     & 0  & 0      & 0  & 0  & 0      & 0      & 0      & 0      \\
16      & 0   & 0 & 0      & 0      & 0   & 0 & 0     & 0 & 0   & 0  & 0     & 0  & 2100   & 0  & 0  & -95400 & 93300  & 0      & 0      \\
17      & 0   & 0 & 0      & 0      & 0   & 0 & 0     & 0 & 0   & 0  & 0     & 0  & 0      & 0  & 0  & 93300  & -93300 & 0      & 0      \\
18      & 0   & 0 & 0      & 0      & 0   & 0 & 0     & 0 & 0   & 0  & 0     & 0  & 2100   & 0  & 0  & 0      & 0      & -95400 & 93300  \\
19      & 0   & 0 & 0      & 0      & 0   & 0 & 0     & 0 & 0   & 0  & 0     & 0  & 0      & 0  & 0  & 0      & 0      & 93300  & -93300\\
\hline\hline
El. & 1   & 2 & 3      & 4      & 5   & 6 & 7     & 8 & 9   & 10 & 11    & 12 & 13     & 14 & 15 & 16     & 17     & 18     & 19     \\\hline
1  & 15000  & 0    & -15000 & 0     & 0      & 0    & 0      & 0     & 0      & 0    & 0      & 0    & 0     & 0 & 0 & 0    & 0     & 0    & 0    \\
2  & 0      & 0.2  & -0.2   & 0     & 0      & 0    & 0      & 0     & 0      & 0    & 0      & 0    & 0     & 0 & 0 & 0    & 0     & 0    & 0    \\
3  & -15000 & -0.2 & 16100  & 0     & 0      & 0    & 0      & 0     & 0      & 0    & 0      & 0    & 11400 & 0 & 0 & 0    & 0     & 0    & 0    \\
4  & 0      & 0    & 0      & 26.3  & 0      & 0    & 0      & 0     & 0      & 0    & 0      & 0    & -26.3 & 0 & 0 & 0    & 0     & 0    & 0    \\
5  & 0      & 0    & 0      & 0     & 15000  & 0    & -15000 & 0     & 0      & 0    & 0      & 0    & 0     & 0 & 0 & 0    & 0     & 0    & 0    \\
6  & 0      & 0    & 0      & 0     & 0      & 0.2  & 0      & 0     & 0      & 0    & 0      & 0    & -0.2  & 0 & 0 & 0    & 0     & 0    & 0    \\
7  & 0      & 0    & 0      & 0     & -15000 & 0    & 15200  & -101  & 0      & 0    & 0      & 0    & -95   & 0 & 0 & 0    & 0     & 0    & 0    \\
8  & 0      & 0    & 0      & 0     & 0      & 0    & -101   & 102   & 0      & 0    & 0      & 0    & -1.48 & 0 & 0 & 0    & 0     & 0    & 0    \\
9  & 0      & 0    & 0      & 0     & 0      & 0    & 0      & 0     & 15000  & 0    & -15000 & 0    & 0     & 0 & 0 & 0    & 0     & 0    & 0    \\
10 & 0      & 0    & 0      & 0     & 0      & 0    & 0      & 0     & 0      & 0.2  & 0      & 0    & -0.2  & 0 & 0 & 0    & 0     & 0    & 0    \\
11 & 0      & 0    & 0      & 0     & 0      & 0    & 0      & 0     & -15000 & 0    & 15200  & -101 & -95   & 0 & 0 & 0    & 0     & 0    & 0    \\
12 & 0      & 0    & 0      & 0     & 0      & 0    & 0      & 0     & 0      & 0    & -101   & 102  & -1.47 & 0 & 0 & 0    & 0     & 0    & 0    \\
13 & 0      & 0    & -1140  & -26.3 & 0      & -0.2 & -95    & -1.48 & 0      & -0.2 & -95    & 1.47 & 32500 & 0 & 0 & 880  & -20.4 & -965 & 24.7 \\
14 & 0      & 0    & 0      & 0     & 0      & 0    & 0      & 0     & 0      & 0    & 0      & 0    & 0     & 0 & 0 & 0    & 0     & 0    & 0    \\
15 & 0      & 0    & 0      & 0     & 0      & 0    & 0      & 0     & 0      & 0    & 0      & 0    & 0     & 0 & 0 & 0    & 0     & 0    & 0    \\
16 & 0      & 0    & 0      & 0     & 0      & 0    & 0      & 0     & 0      & 0    & 0      & 0    & -880  & 0 & 0 & 1980 & -1100 & 0    & 0    \\
17 & 0      & 0    & 0      & 0     & 0      & 0    & 0      & 0     & 0      & 0    & 0      & 0    & -20.4 & 0 & 0 & 1100 & 1120  & 0    & 0    \\
18 & 0      & 0    & 0      & 0     & 0      & 0    & 0      & 0     & 0      & 0    & 0      & 0    & -965  & 0 & 0 & 0    & 0     & 1530 & 562  \\
19 & 0      & 0    & 0      & 0     & 0      & 0    & 0      & 0     & 0      & 0    & 0      & 0    & -24.7 & 0 & 0 & 0    & 0     & -562 & 587  \\
\label{table:table2}
\end{longtable}
\end{landscape}
}

\end{appendix}

\end{document}